\documentclass{caps}
\usepackage{epsfig}

\begin{document}

\pagenumbering{roman}

\pagenumbering{arabic}

\setcounter{chapter}{8}

\author[Rob Fender]{Rob Fender\\Astronomical Institute `Anton
  Pannekoek', University of Amsterdam,\\ Kruislaan 403, 1098 SJ
  Amsterdam, The Netherlands}

\chapter{Jets from X-ray binaries}
\tableofcontents
\section{History}

Relativistic outflows, or `Jets', represent one of the most obvious,
important and yet poorly-explained phenomena associated with accreting
relativistic objects, including X-ray binaries. Originally recognised
in images as long, thin structures apparently connected at one end to
the nuclei of galaxies, it was soon established that they represent
powerful flows of energy and matter away from accreting black holes
and back to the Universe at large. From their earliest association
with the most luminous sources in the Universe, the Active Galactic
Nuclei (AGN), the conclusion could have been drawn that jets were a
common consequence of the process of accretion onto relativistic
objects. Nevertheless, their association with the analogous accretion
processes involving stellar-mass black holes and neutron stars was not
systematically explored until the past decade or so.

Although it is now clear that the electromagnetic radiation from X-ray
binary jets may extend to at least the X-ray band, historically the key
observational aspect of jets is their radio emission. High brightness
temperatures (see section {\ref{physical}}), `nonthermal' spectra and
polarisation measurements indicate an origin as synchrotron emission
from relativistic electrons. Following the discovery of luminous
binary X-ray sources in the 1960s and 1970s, radio counterparts were
associated with the brightest of these, e.g. Sco X-1 (Hjellming \&
Wade 1971a), Cyg X-1 (Hjellming \& Wade 1971b) and the outbursting
source Cyg X-3 (Gregory et al. 1972 et seq.).  However, it was not
until radio observations of the strong radio source associated with
the unusual binary SS 433 revealed a {\em resolved} radio source that
the field of X-ray binary jets really opened up (Spencer 1979; see
also Hjellming \& Johnston 1981a,b). Outbursts of `soft X-ray
transients' were also often associated with strong, transient radio
emission (e.g. A0620-00: Owen et al. 1976; GS 1124-583: Ball et
al. 1995; see also Hjellming \& Han 1995; Kuulkers et al. 1999; Fender
\& Kuulkers 2001).

In the 1990s the study of jets from X-ray binaries entered a new phase
with the discovery of apparent superluminal motions in the outflow
from the bright X-ray transient `microquasar' GRS 1915+105 (Mirabel \&
Rodr\'\i guez 1994; see also Mirabel \& Rodr\'\i guez 1999; Fender et
al. 1999a; Rodr\'\i guez \& Mirabel 1999; Fender et al. 2002). For the
first time it was clear that the jets from X-ray binaries can also
exhibit the kind of significantly relativistic (Lorentz factors
$\Gamma \geq 2$, where $\Gamma = (1-\beta^2)^{-1/2}$ and $\beta =
v/c$) velocities observed in the jets of AGN, and not just the mildly
relativistic velocity of $\sim 0.26c$ ($\Gamma = 1.04$) observed in SS
433. Exactly {\em how} relativistic these jets are will be discussed
later. Shortly afterwards a second superluminal galactic source, GRO
J1655-40, was discovered (Tingay et al. 1995; Hjellming \& Rupen
1995).

Since this period detailed investigations of the jets from X-ray
binaries, both in the radio band and at shorter wavelengths, have
revealed a rich phenomenology and clear patterns of behaviour which
have provided unique insights into the coupling of accretion and
outflow close to relativistic objects. Nevertheless, the deeper we
look the more complex the behaviour becomes, and this is a rapidly
advancing field. In this review I shall attempt, subjectively, to
describe the state of the research as it is at the beginning of 2003.

\begin{figure}
\centerline{\epsfig{file=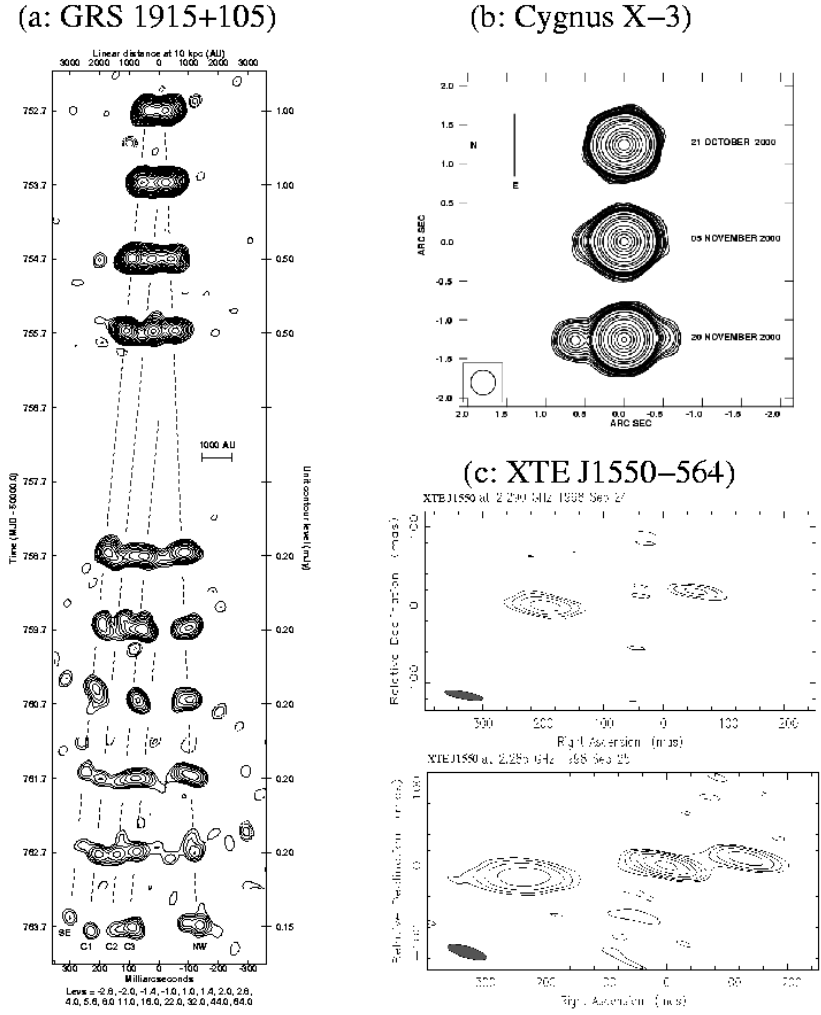,width=13cm,angle=0}}
\caption{Radio images of relativistic jets from X-ray binaries. Panel (a)
  shows a sequence of images of `superluminal' relativistic ejections from
GRS 1915+105 observed with MERLIN (Fender et al. 1999a). Panel (b) is
  a sequence of slower, arcsec-scale jets from Cyg X-3 (Mart\'\i {}  et
  al. 2002), which may be the jet-ISM interaction zones of the inner,
  more relativistic jet (Mioduszewski et al. 2000). Panel (c) presents
  two VLBI images of XTE J1550-564 shortly after the major flare in
  1998 which was probably responsible for the formation of radio and
  X-ray lobes (see Fig {\ref{corbel}}) four years later; from
  Hannikainen et al. (2001).}
\label{jetimages}
\end{figure}

\begin{figure}
\centerline{\epsfig{file=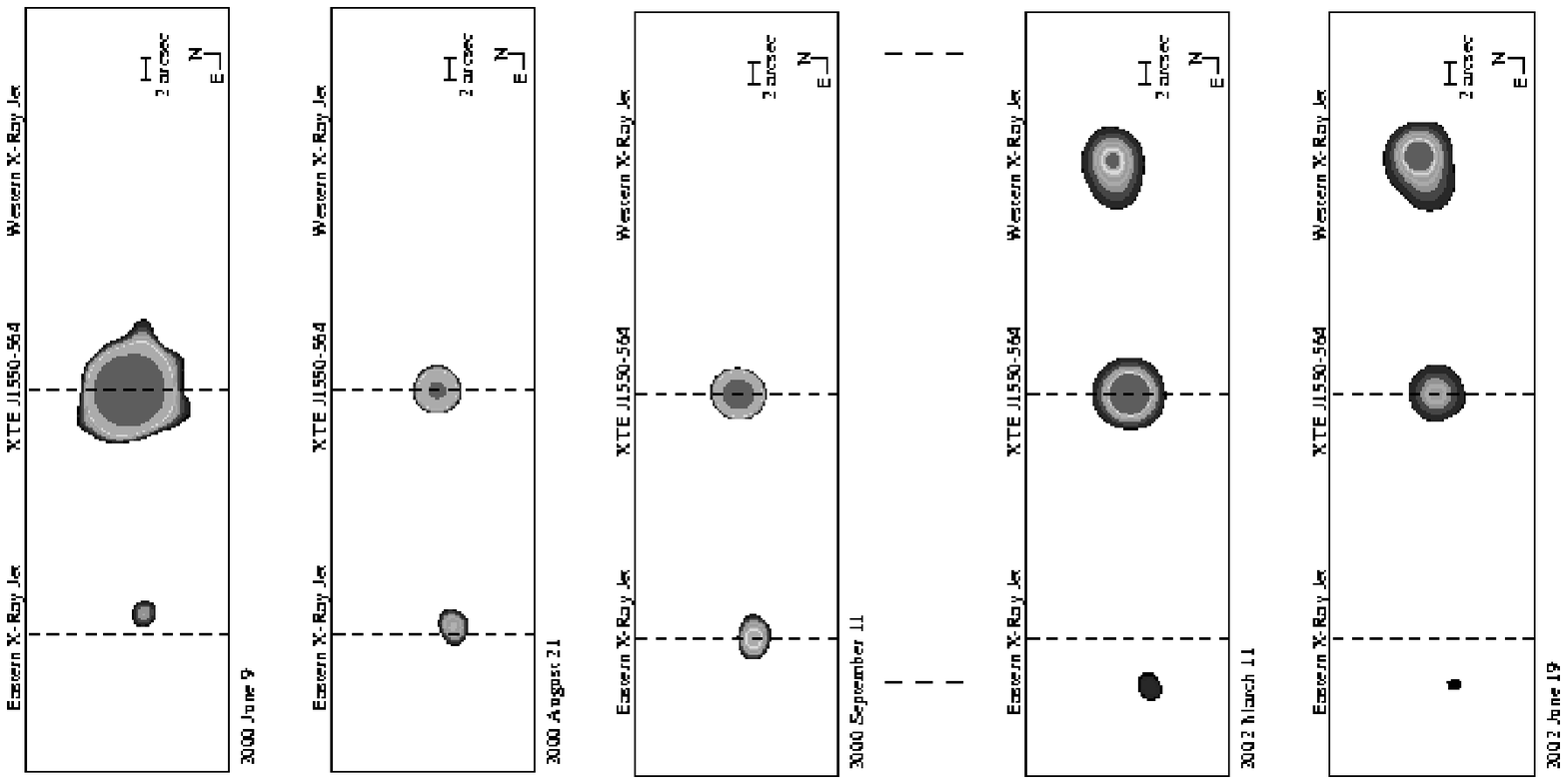,width=17cm,angle=270}}
\caption{{\em Chandra} images of moving X-ray jets from the black hole
transient XTE J1550-564 (Corbel et al. 2002; see also Kaaret et
al. 2002 and Tomsick et al. 2002). The core is the central component,
the `approaching' jet to the left (East) and the `receding' jet to the
right (West). These remarkable observations are the first detections
of relativistic proper motions in X-rays, and demonstrate
unambiguously that X-ray binary jets can accelerate electrons to
extremely high energies and as a result are sources of beamed X-rays.
Note that in the top panel the apparent fluxes are reduced by the
presence of a grating; in fact the eastern jet was brighter at this
epoch than at any time subsequently.
}
\label{corbel}
\end{figure}

In Figs {\ref{jetimages}} and {\ref{corbel}} are presented recent
sequences of observations of transient relativistic outflows from
black hole binaries. Fig {\ref{jetimages}} presents {\em radio} radio
images of relativistic ejections from three outbursting X-ray
binaries on sub-arcsecond angular scales.
Fig {\ref{corbel}} presents {\em X-ray} images of arcsecond-scale
jets moving away from the transient XTE J1550-564 up to four years
after the original ejection event, observed with Chandra.

\section{Physical properties of the jets}
\label{physical}

In the following I shall briefly outline out understanding of the
emission mechanisms in X-ray binary jets, and how we can estimate
important physical quantities from the most basic of observations.

\subsection{Emission mechanism}

The radio jets observed from X-ray binaries emit via the synchrotron
process. We are drawn to this conclusion by their `nonthermal'
spectra, high brightness temperatures and, in some cases, high degree
of linear polarisation. In the following we will illustrate how some
fundamental parameters, e.g. the magnetic field and energy associated
with ejection events, can be estimated from the most basic
observations. For a more detailed explanation and exploration of
synchrotron emission the reader is directed to e.g. Longair (1994).

Bright events associated with e.g. X-ray state changes and X-ray
transients reveal an optically thin spectrum above some frequency,
from which the underlying electron population can be derived. If the
underlying electron distribution is a power law of the form $N(E) dE
\propto E^{-p} dE$ then observations of the spectral index ($\alpha =
\Delta \log S_{\nu} / \Delta \log \nu$, i.e. S$_{\nu} \propto
\nu^{\alpha}$) in the optically thin part of the synchrotron spectrum
can directly reveal the form of this electron distribution: $p =
1-2\alpha$.

Observed optically thin spectral indices $-0.4 \geq \alpha \geq -0.8$,
indicate $1.8 \leq p \leq 2.6$. This is the same range derived for the
majority of AGN jets and also for synchrotron emission observed in
other astrophysical scenarios e.g. supernova remnants, and is
consistent with an origin for the electron distribution in shock
acceleration (e.g. Longair 1994; Gallant 2002).

\subsection{Minimum energy estimation}

Association of a given synchrotron luminosity with a given volume
(either by direct radio imaging or by measurement of an associated
variability timescale) allows estimation of the minimum energy
associated with the synchrotron-emitting plasma (Burbidge 1959), at a
corresponding `equipartition' magnetic field. 


\begin{figure}
\centerline{\epsfig{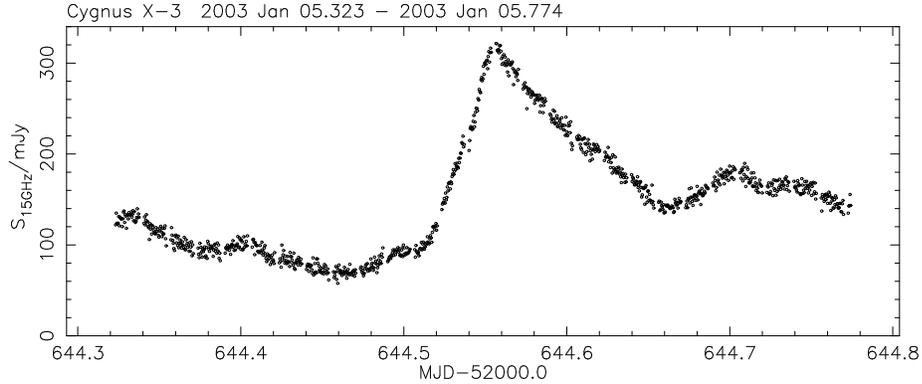}}
\caption{Observation of a `clean' radio flare event from the jet
source Cyg X-3 at 15 GHz. The rise time of the event $\sim 0.04$ d,
allows an estimation of the size of the region associated with the
event, and thus the minimum energy. Observations from the Ryle
Telescope (Guy Pooley, private communication).}
\label{cygx3flare}
\end{figure}

Longair (1994) gives a clear explanation of the calculation of the
minimum energy and corresponding magnetic field, and the interested
reader is directed there. Repeating some of his useful formulae, a
lower limit to the energy associated with a finite volume of
synchrotron emitting plasma can be obtained from a simple estimate of
the monochromatic luminosity at a given frequency which is associated
with that volume:

\begin{equation}
E_{\rm min} \sim 8\times 10^{6} \eta^{4/7} \left(\frac{V}{{\rm
    cm}^3}\right)^{3/7} \left(\frac{\nu}{\rm Hz}\right)^{2/7}
    \left(\frac{L_{\nu}}{{\rm erg \phantom{0}
    s^{-1} Hz^{-1}}}\right)^{4/7} \phantom{00} {\rm erg}
\end{equation}

where $\eta = (1+\beta)$ and $\beta = \epsilon_{\rm p} / \epsilon_{\rm
 e}$ represents the ratio of energy in protons to that in electrons,
and assuming $p=2$.  It is generally accepted that $\beta \sim 0$ and
therefore $\eta \sim 1$, often with little serious justification. In
the more common case where we do not image the source but rather infer
its size from the rise time $\Delta t$ of an event (i.e. using $V =
(4/3) \pi (c \Delta t)^3$ with a flux density $S_{\nu}$ originating at
an estimated distance $d$, the formula can be rewritten as

\begin{equation}
E_{\rm min} \sim 3\times10^{33} \eta^{4/7} \left(\frac{\Delta t}{\rm
  s}\right)^{9/7} \left(\frac{\nu}{\rm GHz}\right)^{2/7}
\left(\frac{S_{\nu}}{\rm mJy}\right)^{4/7} \left(\frac{d}{\rm
  kpc}\right)^{8/7} \phantom{00} {\rm erg}
\end{equation}

The related mean power into the ejection event 

\begin{equation}
P_{\rm min} = \frac{E_{\rm min}}{\Delta t} \sim 3\times10^{33} \eta^{4/7} \left(\frac{\Delta t}{\rm
  s}\right)^{2/7} \left(\frac{\nu}{\rm GHz}\right)^{2/7}
\left(\frac{S_{\nu}}{\rm mJy}\right)^{4/7} \left(\frac{d}{\rm
  kpc}\right)^{8/7} \phantom{00} {\rm erg} \phantom{0} {\rm s^{-1}}
\end{equation}

This minimum energy condition is achieved at so-called
`equipartition', when the energy in particles and magnetic field is
comparable. This field can be approximated by

\begin{equation}
B_{\rm eq} \sim 30 \eta^{2/7} \left(\frac{S_{\nu}}{\rm mJy}\right)^{2/7} \left(\frac{d}{\rm kpc}\right)^{4/7} \left(\frac{\Delta t}{\rm s}\right)^{-6/7} \left(\frac{\nu}{\rm GHz}\right)^{1/7} \phantom{00} {\rm G}
\end{equation}

Note that this field is not, as can sometimes be presumed, a {\em
minimum} magnetic field but rather the field corresponding to the
minimum energy -- i.e. increase or decrease the field and the energy
required to produce the observed synchrotron emission increases. The
Lorentz factors of electrons (or positrons) emitting synchrotron
emission at a given frequency can be estimated by:

\begin{equation}
\gamma_e \sim 30 \left(\frac{\nu}{\rm GHz}\right)^{1/2} \left(\frac{B}{\rm G}\right)^{-1/2}
\end{equation}

Fig {\ref{cygx3flare}} shows a clean radio flare event from the X-ray
binary jet source Cyg X-3. The observation is at 15 GHz, has a rise
time of $\sim 3500$ s, an amplitude of $\sim 200$ mJy and Cyg X-3 lies
at an estimated distance of $\sim 8$ kpc. Using the above
approximations we find a minimum energy associated with the event of
$E_{\rm min} \sim 5\times 10^{40}$ erg, and a corresponding mean jet
power during the event of $\sim 10^{37}$ erg s$^{-1}$, many orders of
magnitude greater than the observed radiative luminosity.
The corresponding equipartition field can be estimated
as $\sim 0.5$ Gauss, in which field electrons radiating at 15 GHz must
have Lorentz factors $\gamma \sim 150$. 

It should be stressed that the inner regions of jets from X-ray
binaries have relativistic Doppler factors (see below) considerably
different from unity resulting from relativistic bulk motions, whereas
the above estimations are based upon rise times, flux densities and
frequencies as measured in the comoving frame. In such cases the
observed quantities need to be corrected to the comoving frame before
the estimates can be made. In addition, in such cases the kinetic
energy associated with the ejection needs to be taken into
account. This kinetic energy component is given by:

\begin{equation}
E_{\rm kin} = (\Gamma-1) E_{\rm int}
\end{equation}

 -- i.e. for a bulk Lorentz factor $\Gamma >2$ (by no means unreasonable --
see below) kinetic dominates over internal energy.

\subsection{Flare events}

Flare events such as that presented in Fig {\ref{cygx3flare}} are
believed to result from the short-term injection of energy and
particles into an expanding plasma cloud, presumably in the form of a
jet. Such events are characterised by optically thin spectra and are
associated with e.g. X-ray transients and persistently flaring sources
such as Cyg X-3 and GRS 1915+105. From Fig {\ref{cygx3flare}} it is
clear that {\em rise} and {\em decay} phases can be quite clearly
defined. In the `synchrotron bubble' model (van der Laan 1966;
Hjellming \& Johnston 1988; Hjellming \& Han 1995 and references
therein) the rise phase corresponds to a decreasing optical depth at
frequencies which were initially (synchrotron-)self-absorbed;
observational characteristics would be an inverted radio spectrum
during the rise phase, and possible Doppler effects on the profile
(since the effect takes place in a different frame to the observer).
An alternative explanation is that the rise phase represents a finite
period of particle injection/acceleration in the outflow; the
characteristics of such a phase would be an optically thin spectrum
and a duration at least coupled to events more or less in the
observer's frame e.g. the X-ray emission arising from the accretion
disc. It seems (to this author) that there are probably observed
events of both types. Note that time delays in the propagation of a
shock (or other particle acceleration phenomena) through the
differing `photospheres' of an outflow may (misleadingly) mimic the
`synchrotron bubble' effect (see discussion in Klein-Wolt et al. 2001).

The monotonic decay observed after a few days in the radio events from
X-ray transients (see below) seems to be primarily due to adiabatic
expansion losses, the key signature of which is the same decay rate at
all frequencies. Significant loss of energy through the synchrotron
emission process itself, or via inverse Compton scattering, results in
a more rapid decay at higher frequencies (spectral steepening). The
fact that adiabatic losses dominate reveals clearly that the
synchrotron radiation observed from such events is only a small
fraction of the total energy originally input.

\subsection{Speed}

Mirabel \& Rodr\'\i guez (1994) first reported apparent superluminal
motions from a galactic source, GRS 1915+105. This apparent velocity
($\beta_{obs}$) is related to the observed proper motion by

\begin{equation}
\beta_{\rm obs} \sim \left(\frac{\mu}{\rm 170 \phantom{0}mas\phantom{0} d^{-1}}\right) \left(\frac{d}{\rm kpc}\right)
\end{equation}

The apparent velocity is related to the intrinsic velocity ($\beta_{int}$)
by:

\begin{equation}
\beta_{\rm obs} = \frac{\beta_{\rm int} \sin \theta}{1 \mp \beta_{\rm int} \cos \theta}
\end{equation}

where $\theta$ is the angle of the flow to the line of sight ($\mp$
refer to approaching and receding components respectively). Apparent
superluminal motion (i.e. $\beta_{obs} > 1$) requires $\beta_{int}
\geq 0.7$, indicating that at least mildly relativistic intrinsic
velocities are required to achieve the effect (or a badly
overestimated distance!). The associated relativistic Doppler shift is
given by

\begin{equation}
\delta = \Gamma^{-1} (1 \mp \beta_{\rm int} \cos \theta)^{-1}
\label{reldopp}
\end{equation}

where $\Gamma$ is the bulk Lorentz factor of the flow. This $\Gamma$ term
represents time dilation at relativistic velocities and means that in
certain circumstances (probably the case for the superluminal jet
sources GRS 1915+105 and GRO J1655-40) {\em both} jets can be
redshifted.

Given observed proper motions of jets, how can we estimate $\beta_{\rm
int}$ ?  As described in Mirabel \& Rodr\'\i guez (1994), measurement
of $\mu_{\rm app}$ and $\mu_{\rm rec}$ allows a determination of the
following product:

\begin{equation}
\beta_{\rm int} \cos \theta = \frac{(\mu_{\rm app}-\mu_{\rm rec})}{(\mu_{\rm
app}+\mu_{\rm rec})}
\end{equation}

where $\theta$ is the angle of the ejection to the line of sight and
$\mu_{\rm app}$, $\mu_{\rm rec}$ are the approaching and receding
proper motions respectively (see also Rees 1966; Blandford, McKee \&
Rees 1977). 

Once the proper motions are measured, the angle of ejection, $\theta$,
and consequently the intrinsic velocity, $\beta$, are uniquely
determined for every distance since

\begin{equation}
\tan \theta = \frac{2d}{c}\left(\frac{\mu_{\rm app}\mu_{\rm
rec}}{\mu_{\rm app}-\mu_{\rm rec}}\right)
\end{equation}

\noindent and the product $\beta_{\rm int} \cos \theta$ is already known.

The variation of $\beta_{\rm int}$ and $\theta$ as a function of
distance for GRS 1915+105 was presented in Fender et al. (1999a). There
is a maximum distance to the source corresponding to $\beta_{\rm int} = 1$
(i.e. $\Gamma = \infty$):

\begin{equation}
d_{\rm max} = \frac{c}{\sqrt(\mu_{\rm app}\mu_{\rm rec})}
\end{equation}

At this upper limit to the distance you also find the maximum
angle of the jet to the line of sight,

\begin{equation}
\theta_{\rm max} = \cos^{-1} \frac{(\mu_{\rm app}-\mu_{\rm rec})}{(\mu_{\rm
app}+\mu_{\rm rec})}
\end{equation}

In addition to the proper motions and Doppler-shifting of frequencies,
there is a boosting effect due to a combination of Doppler
and relativistic aberration effects, both contained in the
relativistic Doppler factor (eqn. {\ref{reldopp}}). An object moving at
angle $\theta$ to the line of sight with velocity $\beta$ (and
resultant Lorentz factor $\Gamma$) will have an observed surface
brightness $\delta^k$ brighter, where $2 < k < 3$ ($k=2$ corresponds
to the average of multiple events in e.g. a continuous jet, $k=3$
corresponds to emission dominated by a singularly evolving
event). Therefore the ratio of flux densities from approaching and
receding knots -- measured at the same angular separation from the
core, so as to sample the knots at the same age in their evolution --
will be given by:

\[
\frac{S_{\rm app}}{S_{\rm rec}} = \left(\frac{\delta_{\rm
app}}{\delta_{\rm rec}}\right)^{k-\alpha}
\]

where $\alpha$ is the spectral index (to compensate for the spectral
shape for different Doppler shifts). For a more detailed discussion
see e.g. Blandford et al. 1977; Hughes 1991; Mirabel \& Rodr\'\i guez
1999; Fender 2003). 

\subsubsection{Observed speeds of steady jets}

There are basically no direct measurements of the speeds associated
with the `steady' jets inferred to exist in the low/hard state of
black holes (see section {\ref{lhssec}}), and possibly also in the
`plateau' state of GRS 1915+105 and the hard states of some neutron
star Atoll sources. Nevertheless, there are some clues that the jets
may be mildly, but not highly, relativistic. Stirling et al. (2001),
in direct imaging of the mas-scale jet from Cyg X-1 in the low/hard
state, inferred a minimum speed of $\beta \geq 0.6$ based upon the
one-sidedness of the jet.  Gallo, Fender \& Pooley (2003) have
performed Monte Carlo simulations in order to investigate the effect
of significant Doppler boosting on the observed radio : X-ray
correlation in the low/hard state (see Fig {\ref{gallo}}). They found
that intrinsic velocities for the radio emitting component of $v >
0.8c$ would probably result in a larger spread in the correlation
than is observed -- therefore the bulk Lorentz factor $\Gamma$ of the
steady radio-emitting jets is likely to be $<2$ (strictly true only
for cases in which the X-rays are not significantly beamed).

It is perhaps incorrect to place the neutron star Z sources in this
section, since they may be just as well considered `persistent
transients', like GRS 1915+105. Whatever the classification, the
observations of Sco X-1 (Fomalont et al. 2001a,b) present a
fascinating demonstration that the velocity of the flow from the
accretion region may be rather different from that observed for the
radio-emitting knots. Specifically, an unseen underlying flow with
Lorentz factor $\geq 2$ is inferred to be powering a particle
acceleration zone which is itself moving away from the binary with a
mildly relativistic (and non-constant) speed of $\sim 0.5c$.

\subsubsection{Observed speeds of transient jets}

In 1994 VLA observations of apparent superluminal motions from the
black hole transient GRS 1915+105 demonstrated unequivocally that
X-ray binaries could produce highly relativistic jets (Mirabel \&
Rodr\'\i guez 1994). Since then, a further three or four superluminal
sources have been discovered (GRO J1655-40: Tingay et al. 1995;
Hjellming \& Rupen 1995; XTE J1748-288: Rupen, Hjellming \&
Mioduszewski 1998; XTE J1550-560: Hannikainen et al. 2001; Corbel et
al. 2002; V4641 Sg: Hjellming et al. 2000a; Orosz et al. 2001), and
there is certainly no indication that highly relativistic ejections
are unusual for black hole X-ray transients.

\begin{figure}
\centerline{\epsfig{file=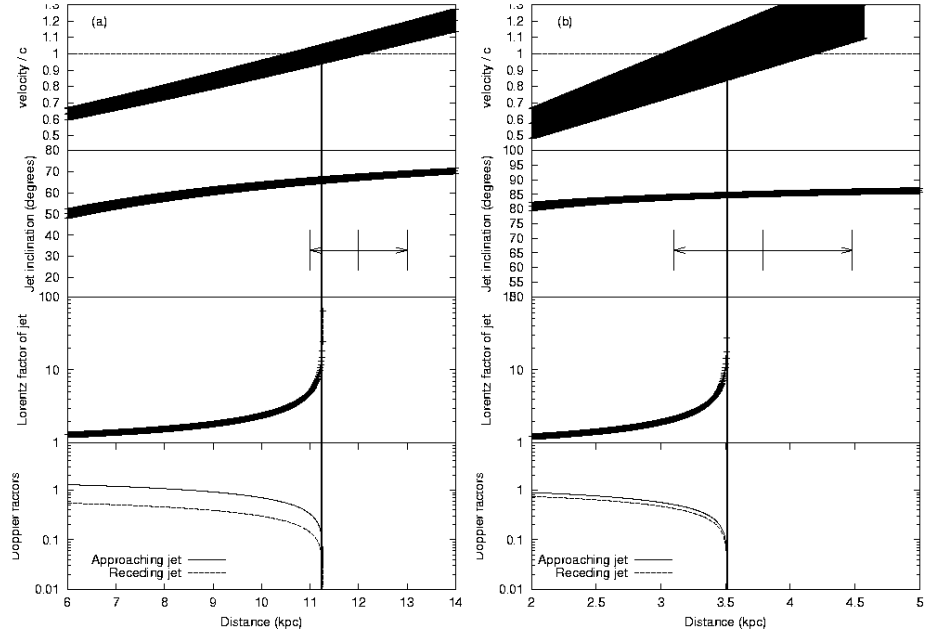,width=10.5cm,angle=270}}
\caption{Variation of solutions to velocity, angle to line of sight,
Lorentz factor and Doppler factors for GRS 1915+105 and GRO J6155-40,
as a function of distance, based upon observations of proper
motions. When compared with the (relatively accurate) distance
estimates it is clear that it is impossible to constrain the Lorentz
or Doppler factors of the flow by such measurements. From Fender (2003).
}
\label{19151655}
\end{figure}

\begin{figure}
\centerline{\epsfig{file=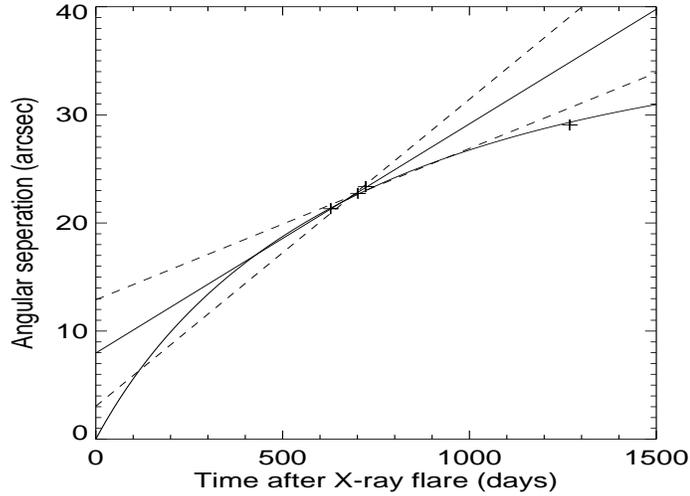,width=9cm,height=6.5cm,angle=0}}
\caption{Deceleration of X-ray jets from XTE J1550-564. Comparing a
  lower limit on the early proper motions on VLBI scales (Hannikainen
  et al. 2001) with subsequent measurements of the X-ray jets (Fig 2)
  with Chandra, indicates a steady slow-down of the jets. A large
  fraction of the dissipated kinetic energy seems to be channeled into
  particle acceleration. From Kaaret et al. (2003); see also Corbel et
  al. (2002) and Tomsick et al. (2003).}
\label{decel}
\end{figure}

But how relativistic are these events ? Following Mirabel \& Rodr\'\i
guez (1994) it was widely accepted that X-ray binary jets could be
characterised by Lorentz factors $\sim 2$ (i.e. while significantly
relativistic, considerably less so than the most extreme examples of
AGN jets). In Fender et al. (1999) it was however shown that a much
wider range of bulk Lorentz factors was possible, at least for GRS
1915+105. Fender (2003) has recently shown that direct measurements of
proper motions of radio components cannot be used to constrain
(specifically, to place an upper limit on) the Lorentz factor of the
flow. In Fig {\ref{19151655}} the solutions to $\beta$, $\theta$,
$\Gamma$ and $\delta_{\rm app, rec}$ are plotted as a function of
distance to the two `superluminal' sources GRS 1915+105 and GRO
J1655-40, along with the best distance estimates. It is clear that
within uncertainties in the distance estimates (which are already
relatively accurate), the Lorentz factor of the jets cannot be
constrained by observations of proper motions.  Nevertheless, Fender
\& Kuulkers (2001) concluded that the mean bulk Lorentz factor for
transients was likely to be $\leq 5$ since higher values would
probably destroy the observed correlation between radio and X-ray peak
fluxes (unless X-ray were also beamed by the same Lorentz factor,
implying inclination selection effects in our source lists). There are
a couple of caveats to this statement: first it has been shown at
least for XTE J1550-564 that jets decelerate steadily as they
propagate away from the binary (Corbel et al. 2002; Kaaret et
al. 2003; see Fig {\ref{decel}}); secondly, the observations of Sco
X-1 (Fomalont et al. 2001a,b) show us that the Lorentz factor (and
hence boosting) of the energising beam may be very different to that
of the actual radio emitting region (consider also V4641 Sgr in this
scenario -- Orosz et al. 2001 and discussion therein).

No proper motions have ever been observed from a neutron-star X-ray
transient. The only concrete hint, physical analogies aside, that they
may be relativistic, is the lower limit of $\geq 0.1c$ for the
arcsec-scale jet of Cir X-1 (Fender et al. 1998) which undergoes a
transient-like outburst every 16.6 days. 

Need the jet velocities be constant ? In SS 433 this seems not to be
the case -- Eikenberry et al. (2002) have shown that the velocity of
the jet may change by more than 10\%. In addition, in XTE J1550-564
(Corbel et al. 2002; Fig {\ref{corbel}}) we clearly observe deceleration
of the jet (Fig {\ref{decel}}). Since this deceleration probably occurs
as a consequence of interactions with the ISM, it is likely to occur
to varying degrees in all X-ray binaries, suggesting that measured
velocities may always be a function of time (a relevant point here is
that there's nothing to indicate that either the original flare event
or the surrounding ISM are particularly unusual in any way).

To summarise, at this stage it seems that the `steady' jets associated
with the low/hard state of black holes and, by analogy, possibly with
some neutron star Atoll sources are only mildly relativistic. The jets
associated with X-ray transients seem almost certain to have
considerably higher Lorentz factors which, however, decrease with time
as the jet interacts with the ISM (see also section
{\ref{interaction}}).  Whether or not there is a smooth continuum of
velocities, or a `switch' from mildly- to highly-relativistic flow
speed (e.g. Meier et al. 1997) is at present unclear.

\subsection{Orientation and precession}

To date it has been assumed, quite reasonably in the absence of other
information, that the jet inclination is perpendicular to the plane of
the binary. However, at least two jet sources (GRO J1655-40 and V4641
Sgr) appear to show significant misalignments (Maccarone 2002 and
references therein).

The clearest example of a precessing jet is SS 433.  The $\sim$162.5-day
precession of these jets (e.g. Margon 1984; Eikenberry et al. 2001)
has been assumed to reflect the precession period of the accretion
disc (see e.g. Ogilivie \& Dubus 2001 for a discussion). Hjellming \&
Rupen (1995) suggested a precession period for GRO J1655-40 which was
very close to the subsequently determined orbital period; similarly
there seems to be marginal evidence for precession in the jets of GRS
1915+105 (Fender et al. 1999; see also Rodr\'\i guez \& Mirabel 1999).
Kaufman Bernado, Romero \& Mirabel (2002) and Romero, Kaufman Bernado
\& Mirabel (2002) have suggested that precessing jets from X-ray
binaries may result in recurrent `microblazar' activity, possibly
manifesting itself as high energy (gamma ray) flashes as the beam
crosses the line of sight. Fender (2003) has discussed the possible
signature of precession on the proper motions observed from a jet
source.

\subsection{Composition}
\label{composition}

Since, with one exception, we have only identified the synchrotron
emission from the leptonic (electrons and/or positrons) component in
X-ray binary jets (a statement also true for AGN), we have little
direct information on their baryonic content (or lack thereof). The
one exception is of course SS 433, whose jets are associated with a
variety of emission lines in optical, infrared and X-ray spectra
(e.g. Margon 1984; Marshall, Canizares \& Schulz 2002).

Why is SS 433 the only jet source with such emission lines ? One
possible interpretation is that all the other jets (which also seem to
have considerably higher bulk velocities than the $\sim 0.26c$
consistently measured for SS 433) have little or no baryonic content
and are dominated by electron : positron pairs. This in turn would
imply that the majority of the mass in the accretion flow never
escapes from the system. It is interesting to note that extended
($\geq$ arcsec) X-ray jets have been observed from both SS 433 and XTE
J1550-564 (Migliari, Fender \& M\`endez 2002; Corbel et al. 2002;
Kaaret et al. 2003; Tomsick et al. 2003; see Fig
{\ref{xspectra}}). The jets from SS 433 reveal strong emission lines
from highly ionised Iron and are consistent with thermal emission from
a plasma at $\sim 10^7$K whereas those from XTE J1550-564 reveal a
featureless continuum which is consistent with an extrapolation of the
synchrotron spectrum from the radio band.  Mirabel et al. (1997) have
discussed effects which would result in atomic emission lines from
significantly relativistic jets being very hard to detect, due to
extreme Doppler broadening in the jet plasma. In addition, Fender
(2003) has shown that the Doppler factors of the jets are very poorly
constrained, so that we basically don't know where to look for such
lines.

An alternative approach to the composition is to investigate the
energetics associated with carrying along a population of `cold'
protons in the relativistic flow. Fender \& Pooley (2000) did this for
the radio--mm--infrared oscillations from GRS 1915+105 and found the
power required to accelerate the proton population to a bulk velocity
$\Gamma = 5$ was so large that the ejections were probably at a
considerably lower bulk Lorentz factor or did not have a large
baryonic component. In a related approach, Celotti \& Ghisellini
(2003) have concluded that a baryonic component is required for the
jets of FRI-type radio galaxies in order to carry most of the power.

An alternative approach to looking for emission lines or balancing
energetics is polarisation -- in particular circular polarisation
holds the promise of a unique insight into the conditions in the
emitting plasma (e.g. Wardle et al. 1998; Wardle \& Homan 2001). Circular
polarisation has been detected from three X-ray binaries -- SS 433
(Fender et al. 2000), GRS 1915+105 (Fender et al. 2002) and GRO
J1655-40 (Macquart et al. 2002). However, the current state of data
and models is not enough to place strong quantitative constraints on
the composition of the jets, since the observed circular polarisation
could arise in both a pair-dominated and baryonic plasma. Right now it
seems that we are no closer to convincingly determining the
composition of jets from X-ray binaries, and the detection of
Doppler-shifted emission (or annihilation) lines from other systems
must remain a high priority observation.

\section{Ubiquity}

While clearly an important physical process for some X-ray binaries,
in order to establish the broader significance of jets from X-ray
binaries it is important to have some idea of their ubiquity. Although
it is always preferable to have directly resolved images of jets, in
many cases it is enough (or at least the best we can do) to infer the
presence of a jet from more circumstantial evidence -- in most cases
this will be e.g. the presence of radio emission with a certain
spectrum or type of variability. This approach can be justified by
considering the following: the (comoving) brightness temperature
$T_{\rm B}$ of an object of physical size $R$, measured with a flux
density $S_{\nu}$ at a frequency $\nu$, and lying at a distance $d$,
is given by the following expression:

\begin{equation}
T_{\rm B} = 2 \times 10^{13} \left(\frac{S_{\nu}}{\rm mJy}\right)
\left(\frac{d}{\rm kpc}\right)^2 \left(\frac{R}{R_{\odot}}\right)^{-2}
\left(\frac{\nu}{\rm GHz}\right)^{-2} \phantom{00} {\rm K}
\end{equation}

Setting a maximum brightness temperature of $T_{\rm B} \leq 10^{12}$
(above which inverse Compton losses become catastrophic, at least for
steady states), this can be rearranged to derive a minimum size for an
emitting region, based upon a measured radio flux density and a
distance estimate.

\begin{equation}
R \geq 4 \left(\frac{S_{\nu}}{\rm mJy}\right)^{1/2} \left(\frac{d}{\rm
  kpc}\right) \left(\frac{\nu}{\rm GHz}\right)^{-1} R_{\odot}
\end{equation}

A typical $\sim 5$ GHz detection of a `weak' radio counterpart to an
X-ray binary is at the $\sim$ mJy level, and such sources typically
lie at distances of $\geq$ 5 kpc. Plugging in those numbers produces a
minimum size for the emitting region $R \geq 8 R_{\odot}$. Typical
binary separations for low mass X-ray binaries are smaller than this;
even the binary separation of Cyg X-1 -- a high-mass X-ray binary in a
relatively large 5.6-day orbit -- is unlikely to be $\geq 15
R_{\odot}$. Therefore we have a relativistic plasma (since the
emission mechanism is synchrotron) with a volume larger than that of
the binary system. Such a plasma will be unconfinable by any known
component of the binary system, and thus will flow out from the
system. Expansion losses will monotonically reduce the flux observed
at optically thin frequencies, and this appears to be the case for the
`synchrotron bubble' events observed from X-ray transients, repeatedly
and clearly resolved by radio interferometers into two-sided outflows.
For the steady sources the same expansion losses require that in order
to observe persistent radio emission, this plasma must be continually
replenished -- therefore we are drawn to conclude that an outflow of
relativistic plasma is present. In nearly all cases, when this
radio-emitting region has been directly resolved, it is in the form of
either steady jet-like structures or outflowing `blobs'; by Occam's
razor we conclude that this is the most likely scenario for most, if
not all, radio emission from X-ray binaries  
(but see Rupen, Mioduszewski \& Hjellming 2002 for the rather
different case of CI Cam).
Note that it is well
known that beamed (ie. relativistically aberrated) emission can
display apparent brightness temperatures $>> 10^{12}$ K, but invoking
relativistic motion to explain away a jet is rather contradictory.
Finally, the same simple jet models originally developed for AGN
naturally reproduce the spectrum and luminosity of radio emission
observed from these systems.

So, allowing ourselves to make the assumption that radio emission is
associated with jets, we can draw the following conclusions, which
will be discussed in greater detail below:

\begin{itemize}
\item{{\em All} black hole systems which are either in the `low/hard'
X-ray state, or are undergoing a major transient outburst, are
associated with the formation of a jet (albeit possibly of different
`types'). Thus the majority of known binary black holes are, or have
been in the past, associated with a jet.}
\item{The six brightest low magnetic field {\em neutron star} systems,
the `Z sources', are all associated with jets in some parts of the `Z'
track. The lower luminosity,
low magnetic field systems, which may be crudely lumped together as
`Atoll' sources, {\em may} be associated with radio emission (although
as with black holes there may be bright soft states without jets),
implying that the lack of radio detections of the majority is a
sensitivity issue.}
\item{The high magnetic field neutron stars, including all but two of
the accreting X-ray pulsars, are {\em not} associated with radio
emission}
\end{itemize}

Adding up the numbers, this author concludes that the evidence for a
jet is very strong in about thirty X-ray binaries (10--15\% of the
currently known population), {\em but} that it is rather likely that
jets are present in up to 70\% of the systems (basically all except
the high magnetic field X-ray pulsars, and a small number of black
hole and neutron star systems which are in persistent `soft' states). 

\section{Disc--jet coupling in black hole binaries}


One of the richest areas of X-ray binary jets research in the past few
years has been the disc:jet coupling, i.e. the relation between inflow
and outflow. Some early clues to the phenomenology outlined below were
reported earlier in the literature -- e.g. some low/hard state
transients were known to exhibit flat-spectrum `second stage' radio
emission (Hjellming \& Han 1995 and references therein) which we would
now associate with the compact jet in the core.  Furthermore,
McCollough et al. (1999) already reported the bimodal behaviour of the
radio:X-ray correlations in Cyg X-3, undoubtedly related to the
changing disc:jet coupling outlined below.

Black holes exhibit, broadly speaking, several different kinds of
X-ray `state'. The two most diametrically opposed, which serve to
illustrate the relation of jet formation to accretion, can be briefly
summarised as:

\begin{itemize}
\item{{\bf Low/Hard (and `Off') state:} in this state the X-ray
spectrum is dominated by a broadband component which can be fit with a
power-law of photon index $\sim 1.6$, often with a cut-off around 100
keV. Minor additional components to the X-ray spectrum include
(sometimes) a weak `black body' (accretion disc) component, a
`reflection' component and a relatively weak gamma-ray tail. The X-ray
power
spectra show up to 40\% r.m.s. variability with a `break' at
frequencies of around a few Hz.}
\item{{\bf `High/Soft' state:} in this state the X-ray spectrum is dominated
by a `black body' component with a temperature around a few keV, with
additional line features and a relatively strong gamma-ray tail. The
X-ray power spectrum shows much less variability, and can be
characterised by a power-law with an r.m.s. variability of only a few
\%.}
\end{itemize}

\begin{figure}
\centerline{\epsfig{file=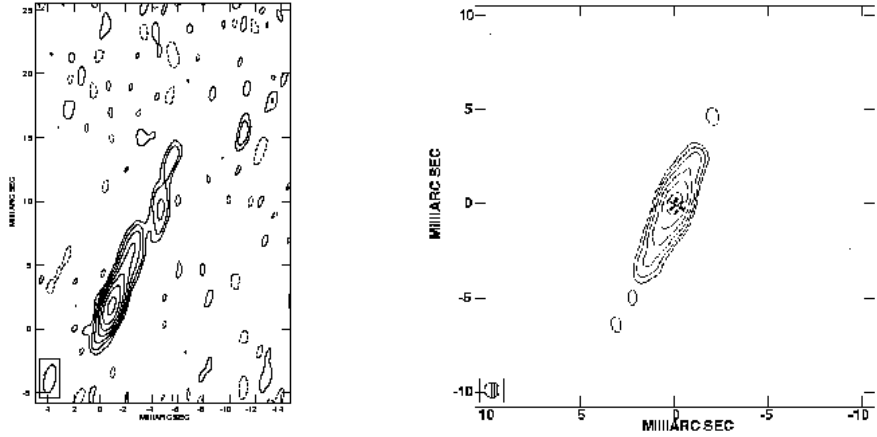,width=12cm,angle=0}}
\caption{AU-scale jets in persistent hard X-ray states, imaged with
the VLBA. The left panel
reveals a one-sided jet from Cygnus X-1 in the classical `low/hard'
X-ray state (Stirling et al. 2001). The right panel shows the
quasi-steady jet from GRS 1915+105 in hard `plateau' states (Dhawan,
Mirabel \& Rodr\'\i guez 2000).}
\label{lowhard}
\end{figure}

Further details of black hole states may be found in the chapters by
van der Klis (chapter 2) , and McClintock \& Remillard (chapter 4).

There are also `Intermediate' and `Very High' states which actually
both appear to be quasi-steady states which share some of the
characteristics of both of the above states, but -- crucially for
their relation to jet formation -- which are {\em much
softer} than the regular `Low/Hard' state. Homan et al. (2001) have
shown that such states can actually occur at a wide variety of
luminosities. 


\begin{figure}
\centerline{\epsfig{file=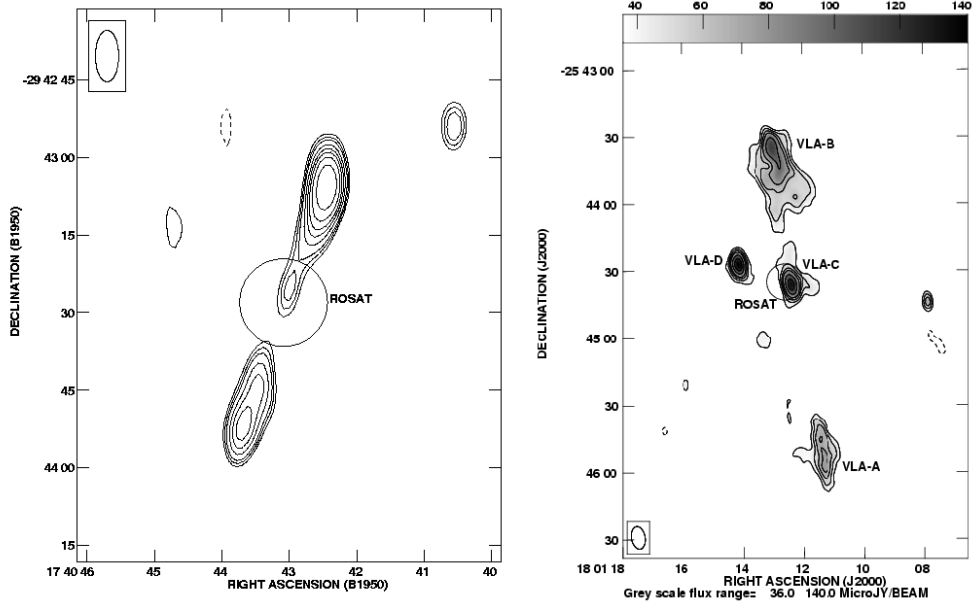,width=11cm,angle=270}}
\caption{Arcmin-scale radio jets from the galactic centre low/hard
state sources 1E 1740.7-2942 and GRS 1758-258. Both of these systems
spend nearly all their time in the low/hard X-ray state, therefore an
interpretation of these lobes is that they result from the long-term
action of steady jets on the ISM. From Mirabel et
al. (1992) and Mart\'\i {}  et al. (2002).}
\label{17401758}
\end{figure}

\subsection{Steady jets in `low/hard' and `quiescent' states}
\label{lhssec}

The radio, and hence jet, properties of the low/hard state black holes
can be summarised thus: a `flat' spectrum (spectral index $\alpha \sim
0$) extending through and beyond the radio band, linear polarisation
at a level of $\sim$ 1--3\% and variability correlated with the X-ray
flux. These broad properties, significantly different from those
associated with transient ejection events, are found in every low/hard
state source (Fender 2001 and references therein).  By analogy with
AGN, it was already suggested that these properties could be explained
by a compact, self-absorbed jet (Hjellming \& Johnston 1988; Falcke \&
Biermann 1996, 1999; Fender 2001; see also Blandford \& K\"onigl
1979). Recently this interpretation has been confirmed by direct
imaging of a milliarcsecond-scale jet from Cyg X-1 in the low/hard
state (Fig {\ref{lowhard}} (left), Stirling et al. 2001); by analogy
it is argued that all low/hard state sources are producing jets.

Furthermore, the hard `plateau' state in GRS 1915+105, which has many
similarities to the classical low/hard state, is also associated with
a resolved milliarsec-scale jet (Dhawan, Mirabel \& Rodr\'\i guez
2000; Fig {\ref{lowhard}} right panel), and the two galactic centre
low/hard state sources 1E 1740.7-2942 \& GRS 1758-258 are both
associated with large-scale radio lobes, indicating the long-term
action of a jet on the local ISM (Mirabel et al. 1992; Mart\'\i {} et
al. 2002; Fig {\ref{17401758}}).

\subsubsection{Spectral extent and jet power}

The radio spectrum in the low/hard state is `flat' or `inverted', in
the sense that the spectral index $\alpha \geq 0$. This spectral
component has been shown to extend to the mm regime for two low/hard
state sources, Cyg X-1 and XTE J1118+480 (Fender et al. 2000b; Fender
et al. 2001). In Fender (2001) it was suggested that correlated
radio--optical (and in fact X-ray) behaviour in the low/hard state
transient V404 Cyg might suggest an extension of the jet spectral
component to the infrared or optical bands. In fact in most, maybe
all, low/hard state sources the optical flux densities seem to lie on
a rather flat ($\alpha \sim 0$) extension of the radio(--mm) spectrum
(e.g. Brocksopp et al. 2001; Corbel et al. 2001).  Jain et al. (2001)
have observed a secondary maximum in the near-infrared light curve of
XTE J1550-564 corresponding to a transition to the low/hard state,
which they also attribute to synchrotron emission from a jet. Rapid
optical variability from XTE J1118+480 in the low/hard X-ray state has
also been interpreted as (cyclo-)synchrotron emission (Merloni, Di
Matteo \& Fabian 2000; see also Hynes et al. 2003) and may be
associated with a sub-relativistic outflow (Kanbach et al. 2001;
Spruit \& Kanbach 2002).

Note that while admittedly not a canonical low/hard state source,
there is unambiguous evidence for synchrotron emission from the jet
source GRS 1915+195 extending at least to the near-infrared band
(Fender et al. 1997; Mirabel et al. 1998; Eikenberry et al. 1998a,
2002; Fender \& Pooley 1998, 2000). Furthermore, but not well
explained, is the correlation in this source between infrared line
strength and synchrotron continuum (Eikenberry et al. 1998b),
indicating a coupling between thermal and nonthermal components.
Qualitatively similar infrared flares have been observed from Cyg X-3
(e.g. Mason, Cordova \& White 1986; Fender et al. 1996) which with the
benefit of hindsight seem likely to be synchrotron in origin. Finally,
Sams, Eckart \& Sunyaev (1996) have observed {\em extended} infrared
emission from GRS 1915+105 which they suggest originates in a jet
(while possibly treated with some scepticism at the time, the
observation of considerably larger X-ray jets from XTE J1550-564 makes
a jet origin seem entirely plausible).

If the flat/inverted radio spectrum is due to self-absorbed synchrotron
emission from a conical jet (Blandford \& K\"onigl 1979; Hjellming \&
Johnston 1988) then above some frequency (at which point the whole jet
is optically thin) there should be a break to an optically thin
spectrum with $-1 \leq \alpha \leq 0$. A compilation of observations
of the low/hard state source GX 339-4 appears to have identified just
such a cut off in the near-infrared (Corbel \& Fender 2002).

How do we estimate the power associated with this steady,
self-absorbed, synchrotron component ?  Without large amplitude
variability, or directly resolved jets, it is not possible to
associate a given luminosity with a certain volume, and it is not
possible to directly apply standard `minimum energy' arguments (as
outlined in section 9.2).  Therefore we must apply other arguments in
order to estimate the total jet power.  In this case we may estimate
the total jet power by (a) carefully measuring the extent of the
synchrotron spectrum which it produces, and (b) introducing a
radiative efficiency, $\eta$, which is the ratio of total to radiated
power (in the jet's rest frame). From this we can estimate the jet
power as

\begin{equation}
P_{\rm J} \sim L_{\rm J} \eta^{-1} F(\Gamma, i)
\end{equation} 

where $L_{\rm J}$ is the total radiative luminosity of the jet
(i.e. the integral of $L_{\nu}$ to the highest measured frequency),
$\eta$ is the radiative efficiency, and $F(\Gamma, i)$ is a correction
factor for bulk relativistic motion with Lorentz factor $\Gamma$ and
Doppler factor $\delta$, ($F(\Gamma, i) \sim \Gamma \delta^{-3}$ --
see Fender 2001).

\begin{figure}
\centerline{\epsfig{file=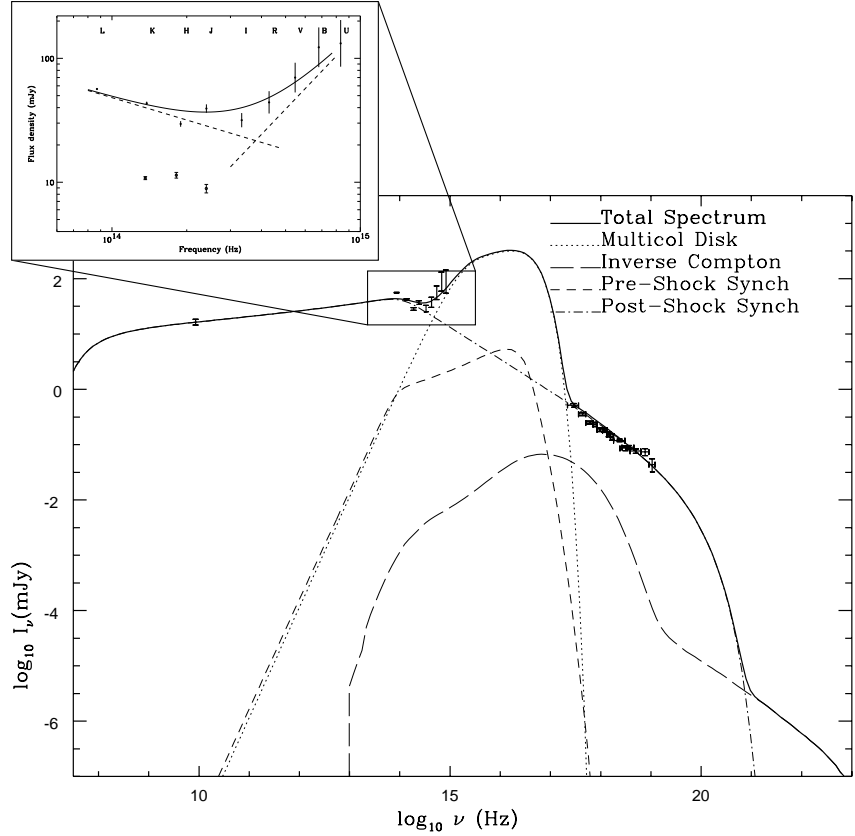,width=13cm,angle=270}}
\caption{Broadband jet-model fit to the radio--X-ray spectrum of GX
  339-4 in the low/hard X-ray state (Markoff et al. 2003). The flat
  spectrum, self-absorbed, synchrotron component extends beyond the
  radio band and breaks to optically thin emission in the
  near-infrared (see insert, from Corbel \& Fender 2002). An
  extrapolation of this near-infrared emission connects smoothly to
  the X-ray power law, suggesting that it may also be optically thin
  synchrotron emission, contrary to more widely-accepted
  Comptonisation models. The broadband spectrum and model fit are
  comparable to those for XTE J1118+480 while in the same X-ray state
  (Markoff, Falcke \& Fender 2001).}
\label{mff}
\end{figure}

Starting from the reasonable assumption that all the emission observed
in the radio band is synchrotron in origin, we can try to see how far
this spectrum extends to other wavelengths. Firstly, it should be made
clear that most systems have not been observed at $\nu < 1$ GHz
(although it appears that the flat radio spectrum of Cyg X-1 extends
at least as low as 350 MHz -- de Bruyn, private communication), and
while some low-frequency turnovers may have occasionally been observed,
there are no reported cases of a complete cut-off to the synchrotron
emission at low radio frequencies. In any case, while a low-frequency
cut-off is important for estimating the mass of the ejecta in the (by
no means certain) case that there is a proton for each emitting
electron, the radiative luminosity is dominated by the high-frequency
extent of the synchrotron spectrum. 

Possibly the most comprehensive broadband spectrum compiled for a
low/hard state source is that for the transient XTE J1118+480, which
clearly shows excess emission at near-infrared and probably also
optical wavelengths (Hynes et al. 2000) and whose radio spectrum
smoothly connects to a sub-mm detection at 850 $\mu$m (Fender et
al. 2001). In Fender et al. (2001) it is argued that in this case the
synchrotron radiative luminosity is already $\geq 1$\% of the
bolometric X-ray luminosity. How important the total jet power is then
depends on our estimates for the radiative efficiency, $\eta$. 


In Fender \& Pooley (2000) an estimate of $\eta$ was made for the
radio `oscillation' events from GRS 1915+105, and an upper limit of
$\eta \leq 0.15$ obtained.  In the original model of Blandford \&
K\"onigl (1979), it is likely that $\eta \leq 0.15$. In the model of
Markoff, Falcke \& Fender (2001; specifically for XTE J1118+480) $\eta
<0.1$. Finally it should be noted that Celotti \& Ghisellini (2003)
estimate $\eta \leq 0.15$ for a sample of AGN. In reality, for the
synchrotron process in jets it seems unlikely theoretically that $\eta
> 0.2$, and this is backed up by no observational counter evidence.
Therefore, for XTE J1118+480 the power in the jet is likely to be
$\geq 10$\% of the X-ray luminosity. Since all low/hard state sources
show a similar broadband spectrum (excluding the influence of
different types of mass donor which only affects the near--infrared
and optical bands) we're drawn to the conclusion that all low/hard
state sources produce powerful jets (Fender 2001).


\subsubsection{Coupling to X-ray emission}

A broad correlation between the radio and X-ray fluxes from a black
hole binary in a low/hard state was first noted by Hannikainen et
al. (1998) for GX 339-4. A similar correlation between radio and X-ray
fluxes was found for Cyg X-1 (Brocksopp et al. 1999), and Fender
(2001) suggested that the magnitude of the radio:X-ray flux ratio was
similar for all low/hard state black holes.

\begin{figure}
\centerline{\epsfig{file=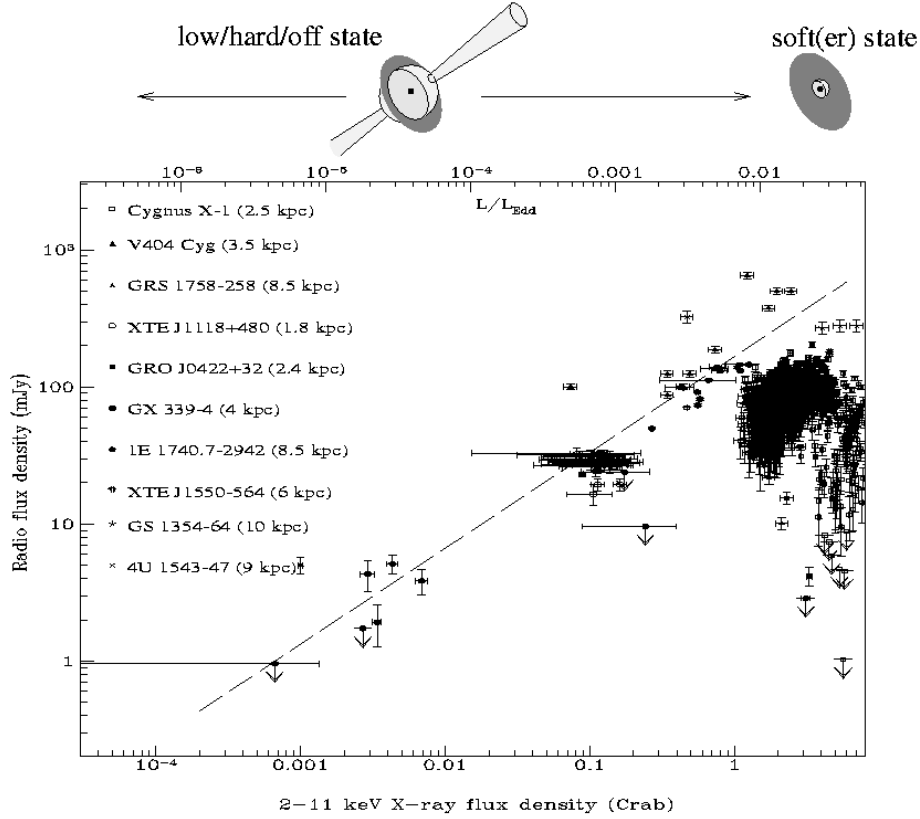,width=12.5cm,angle=270}}
\caption{(Quasi-)simultaneous radio and X-ray observations of black
  hole X-ray binaries, scaled to 1 kpc and corrected for
  absorption. Below a scaled X-ray flux of a few Crab (corresponding
  to $\sim 1$\% of the Eddington luminosity for a 10 $M_{\odot}$ black
  hole), all black hole binaries follow a correlation of the form
  $S_{\rm radio} \propto S_{\rm X-ray}^{0.7}$, in the `low/hard' and
  `off/quiescent' states. The relatively narrow distribution of data
  around a best-fit relation requires that the bulk Lorentz factor of
  jets in the `low/hard' state $\Gamma < 2$. At higher luminosities in
  the (relatively rare) `high/soft' state the radio emission is
  strongly suppressed. At still higher luminosities, X-ray transients
  (including recurrent sources such as Cyg X-3 and GRS 1915+105)
  produce repeated bright optically thin ejections. The hard `plateau'
  state of GRS 1915+105 lies on an extension of the `low/hard' state
  coupling. From Gallo, Fender \& Pooley (2003); see also Corbel et
  al. (2001, 2003).}
\label{gallo}
\end{figure}

In the past couple of years our understanding of this coupling between
radio and X-ray emission has advanced significantly. Corbel et
al. (2000, 2002), in a detailed long-term study of GX 339-4, has found
that the radio emission in the low/hard state scales as $L_{\rm radio}
\propto L_{\rm X-ray}^b$ where $b \sim 0.7$ for X-rays up to at least
20 keV (possibly steepening towards at linear relationship at the
highest X-ray energies). This relation holds over more than three
orders of magnitude in soft X-ray flux. 

\begin{figure}
\centerline{\epsfig{file=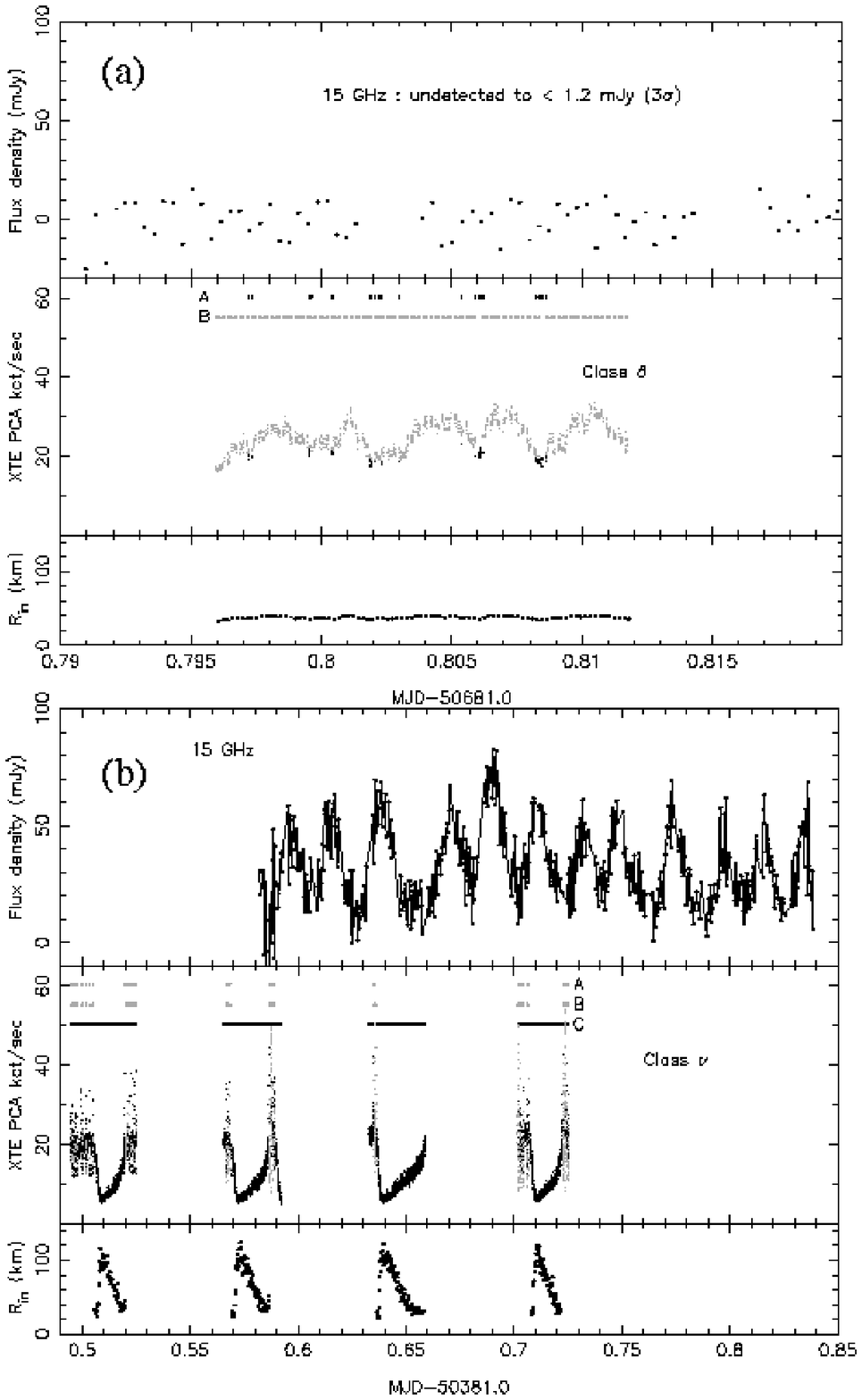,width=10cm,height=11cm,angle=0}}
\caption{GRS 1915+105 often cycles repeatedly between three X-ray
  states: A and B are disc-dominated and `soft'; C is much harder
  (Belloni et al. 2000). Panel (a) above shows that when the source
  only exhibits soft states A and B, the radio emission is very weak;
  however when state C is present the radio emission is much stronger
  (and in fact there is a one-to-one correspondence between state C
  `dips' and radio oscillation events). From Klein-Wolt et al. (2002).}
\label{onoff}
\end{figure}

More recently Gallo, Fender \& Pooley (2002, 2003) have found almost
exactly the same coupling (in both normalisation and slope) over a
comparable range in X-ray luminosity, for the low/hard state transient
V404 Cyg. Furthermore, by compiling data for ten low/hard state
sources, it was found that in the luminosity range $10^{-5} L_{\rm
Edd} \leq L_{\rm X} \leq 10^{-2}L_{\rm Edd}$ all systems are
consistent with the same coupling with a very small scatter (less than
one order of magnitude in radio flux), and
that above a few \% of $L_{\rm Edd}$ the radio emission is rapidly
`quenched' (Gallo et al. 2003). Monte-Carlo simulations of
Doppler-boosting effects indicate that such a small spread over such a
large range in $L_{\rm X}$ probably restricts the velocity of the jet
in the low/hard state to $\beta = v/c \leq 0.8$ (Gallo et al. 2003),
unless the X-rays are also strongly beamed (in which case strong
selection effects are at work).

\subsubsection{Jets in `quiescence' ?}

Outside of periods of transient outburst, BHCs are typically observed
with luminosities in the range $10^{-6}$--$10^{-9}$ Eddington, and are
considered to be `quiescent' (e.g. Garcia et al. 2001). Their X-ray
spectra are generally not distinguishable from the `low/hard' state
however, suggesting that they may also be associated with (relatively)
powerful jets. In fact, V404 Cyg -- the most luminous `quiescent'
black hole -- is clearly associated with a relatively bright and
variable radio source (e.g. Hjellming et al. 2000b) and GX 339-4
follows the radio -- X-ray coupling discussed above down to comparable
X-ray luminosities. In fact, combining the estimates of jet power in
the low/hard state with the $L_{\rm radio} \propto L_{\rm
X-ray}^{0.7}$ relation indicates that `quiescent' BHCs will in fact be
`jet-dominated', in the sense that most of the power output will be in
the form of an outflow (Fender, Gallo \& Jonker 2003). Combining this
result with the greater `radio loudness' of BHCs compared to NS X-ray
binaries can furthermore explain the discrepancy in their `quiescent'
X-ray luminosities (it is observed that NS transients are brighter
X-ray sources in quiescence) without any significant advection of
accretion power across a black hole event horizon (Fender, Gallo \&
Jonker 2003; see also Campana \& Stella 2000; Garcia et al. 2001;
Abramowicz, Kluzniak \& Lasota 2002 and references therein for a
broader discussion of this controversial issue).

\subsection{Loss of jet in `high/soft' states}

The first indication that radio jets are not associated with `soft'
X-ray states can be traced back to Tananbaum et al. (1972), in which
the appearance of the radio counterpart of Cyg X-1 was associated with
a transition from the soft state back to the hard state (see also
Hjellming, Gibson \& Owen 1975). However, while it was surmised that
changes in radio emission were associated with changes in the X-ray
`state' of X-ray binaries (Hjellming \& Han 1995 and references
therein), no clear pattern was established (except perhaps in Cyg X-3,
where is has been realised for some years that periods of `quenched'
radio emission generally preceded large radio outbursts --
e.g. Waltman et al. 1996).

The situation changed when GX 339-4 spent a year in the high/soft
X-ray state in 1998. Radio monitoring of the source in the low/hard
state prior to 1998 had already established the existence of a weak,
mildly variable radio counterpart (Hannikainen et al. 1998), but
throughout the soft state no radio counterpart was detected, despite
multiple observations (Fender et al. 1999b). The source subsequently
returned to the low/hard X-ray state and the weak radio counterpart
reappeared (Corbel et al. 2000). Here was the strongest evidence that
in `soft' disc-dominated states the radio jet was either non-existent
or more than an order of magnitude weaker.

Comprehensive radio and X-ray monitoring of Cyg X-1 has revealed that
the suppression of the radio emission occurs rather rapidly once the
transition to the high/soft state occurs at a bolometric luminosity
of a few per cent Eddington
(Gallo, Fender \& Pooley 2003; Maccarone
2003). Given that there are no observed counter examples, we conclude
that the soft X-ray state does is never associated with a strong radio
jet.  This assertion is supported by the detailed studies of GRS
1915+105 reported by Klein-Wolt et al. (2002), in which steady `soft'
X-ray states are never associated with bright radio emission (Fig
{\ref{onoff}}). 

\begin{figure}
\centerline{\epsfig{file=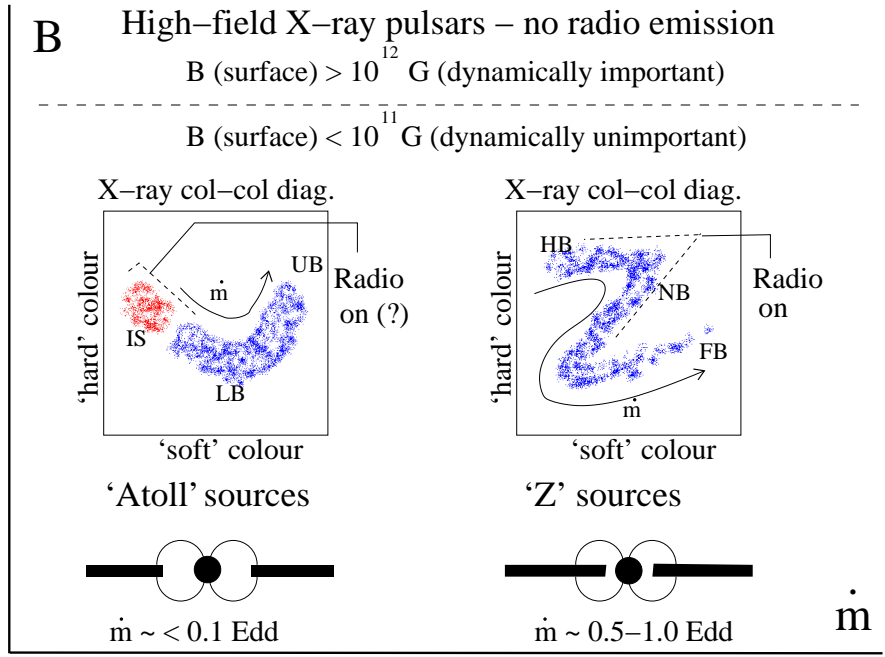,width=10cm,angle=270}}
\caption{Schematic illustrating our current understanding of the
  relation between radio emission and X-ray state for the persistent
  neutron star X-ray binaries}
\label{ns4bw}
\end{figure}

\subsection{`Intermediate' and `Very high' X-ray states}

While the `low/hard' and `high/soft' X-ray states appear to represent
both the most diametrically opposed and the most stable of accretion
modes associated with black hole XRBs, there are also hybrid
states. Both the `intermediate' and `very high' (see McClintock \&
Remillard in this book for more details) states are intermediate in
their X-ray hardness between the two aformentioned canonical
extremes. It has been suggested that they are the same state, in which
case it is a curious fact that this state can occur over quite a large
range in bolometric X-ray luminosity (see Homan et al. 2001).

Belloni (1998) suggested that the behaviour of GRS 1915+105,
oscillating between relatively hard and (two) soft states (see also
Belloni et al. 2000) was reminiscent of the `very high' state as
observed from other luminous X-ray transients. Since these oscillation
events are unambiguously correlated with radio flaring (e.g. Pooley \&
Fender 1997; Mirabel et al. 1998; Klein-Wolt et al. 2002 -- see Fig
{\ref{onoff}}), a connection was made between this state and episodic
jet production.

However, in a very important observation, Corbel et al. (2001) have
shown that in a transition from the low/hard state to the intermediate
state, the radio emission from XTE J1550-564 was reduced by a factor
$>50$. Furthermore, the state in which the jet from Cyg X-1 is
suppressed (see Fig {\ref{gallo}}) may not be the canonical
`high/soft' state, but the `intermediate' state (e.g. Belloni et
al. 1996; Miller et al. 2002a, but see Gierlinski et al. 1999). What
remains clear is that when the X-ray spectrum softens the jet is
suppressed. What needs further investigation is the exact evolution of
the X-ray spectral and jet parameters as this suppression occur, since
at present the most comprehensive studies (e.g. Corbel et al. 2003;
Gallo et al. 2003) are based only on X-ray flux, not spectral,
evolution. In a related work, Pottschmidt et al. (2000) report that
the magnitude of X-ray time lags in Cyg X-1 is much greater {\em
during} transitions than either before or after, and suggest that this
effect may be related to the formation of outflows at these times.

\subsection{The highest luminosities and X-ray transients}

X-ray transients typically peak at luminosities greater than those
which generally characterise the `high/soft' state, although often
still sub-Eddington (Chen, Shrader \& Livio 1997). Such high
luminosities are, in nearly all cases, very short lived (typically
days or less) and the `state' is considerably more difficult to
characterise than the canonical `low/hard' or `high/soft' states. 

It was already known since the 1970s that bright X-ray transients were
associated with transient production of radio emission, whose
characteristics could be described at a basic level by `synchrotron
bubble' models (Hjellming \& Han 1995 and references therein, and also
section 9.2.3).  In several, perhaps all, cases, there is evidence for
{\em multiple} ejection events (e.g. Harmon et al. 1995; Kuulkers et
al. 1999; Brocksopp et al. 2001). The clearest difference with
low/hard state steady jets is the rapid evolution to an optically thin
spectrum ($\alpha \sim -0.6$) and monotonic decay (Fender 2001).  In
addition, linear polarisations of up to a few $\times 10$\% have been
measured (e.g. Fender et al. 1999a; Hannikainen et al. 2000), and also
circular polarisation at the $\sim 1$\% level (see section
{\ref{composition}}). The broad properties of these transient radio
events -- ie. the spectral evolution {\em and} a tendency for multiple
ejection events -- seem to be similar whether the events are `rare'
(e.g. A0620-00, GS 1124-68) or `frequent' (e.g. Cyg X-3, GRS 1915+105).

These ejection events appear to be associated with the change in X-ray
state between `off' (which may be analogous to `low/hard') and very
bright 'high/soft' or `very high' states (e.g. Harmon et al. 1995;
Fender \& Kuulkers 2001). Some transients actually
seem only transit to bright `low/hard' states and may (e.g. V404 Cyg) or
may not (e.g. XTE J1118+480) also display bright optically thin events.
One source seems to sit persistently at close to Eddington
luminosities, and is a spectacular source of relativistic jets: GRS
1915+105. This source exhibits a wide range of X-ray properties, none
of which can be easily classified as normal `low/hard' or `high/soft'
states (Belloni et al. 2000). Its overall X-ray properties may be
reminiscent of the `very high state' (Belloni 1998), but the erratic
flips between hard and (two sorts of) soft states is rather unlike any
other X-ray binary. However, GRS 1915+105 does fit into the general
pattern associating `hard' X-ray states with jet formation, at least
for the `plateau' and `oscillation' events (Dhawan et al. 2000;
Klein-Wolt et al. 2002). Mirabel et al. (1998) have suggested that a
brief X-ray spike, during which the source X-ray spectrum softens
considerably, may indicate the `launch moment' of the jet -- this
would clearly be an important discovery if true and merits further
attention. Cyg X-3 may be displaying similar behaviour to GRS 1915+105
-- it is certainly accreting at a very high level and almost
continuously producing jets -- but details of its workings are hidden
in the dense wind of its Wolf-Rayet companion.

These radio flares (see section 9.2.3) have by now been clearly and
repeatedly associated with highly relativistic bulks motions (section
9.2.4). In a comparison of peak radio and X-ray emission from
transients, Fender \& Kuulkers (2001) found that there appears to be
nothing special about the sources in which relativistic jets had been
resolved. Therefore it seems reasonable to assume (Occam's razor) that
the initial radio emission associated with X-ray transients is always
associated with a relativistic outflow. Note also that Garcia et
al. (2003) have suggested that the largest-scale resolved radio jets
may be associated with X-ray transients with the relatively long
orbital periods.

\section{Disc-jet coupling in neutron star binaries}

As noted in the introduction, radio emission seems to be associated
with both `Z' and `atoll' type neutron star X-ray binaries, but not
with the high-field X-ray pulsars. See Fig {\ref{ns4bw}} for a summary
of our current understanding. It is interesting as a historical note
that a predictable coupling between X-ray `state' and radio emission
was first suggested for the Z sources (see below), but that in recent
years nearly all the attention has switched to the analogous coupling
in black hole systems.

\subsection{`Z' sources}

The prototype Z source, Sco X-1, has been known as a variable radio
source since the early 1970s (Hjellming \& Wade 1971b). This source,
together with GX 5-1, GX 17+2, GX 349+2, GX 340+0 and Cyg X-2 form a
group of neutron star X-ray binaries accreting at or near to the
Eddington limit and exhibiting clear patterns of spectral and timing
behaviour (Hasinger \& van der Klis 1989). It is ironic that an
initial association with large-scale radio lobes was disproved by the
same proper motion studies (Fomalont \& Geldzahler 1991) which
subsequently discovered highly-relativistic jets on milliarcsecond
scales (Bradshaw, Fomalont \& Geldzahler 1999; Fomalont et
al. 2001a,b).  The other five Z sources also have radio counterparts
with comparable luminosities (Penninx 1989; Hjellming \& Han 1995;
Fender \& Hendry 2000).

Priedhorsky et al. (1986) first suggested that an empirical coupling
between X-ray and radio (and optical) emission existed for Sco X-1.
Penninx et al. (1988) confirmed and refined this pattern of behaviour
for GX 17+2 and Penninx (1989) suggested that all Z sources would
display comparable behaviour. The same pattern of behaviour has been
established for Cyg X-2 (Hjellming et al. 1990a) but apparently not in
GX 5-1 (Tan et al. 1991; but see below for a possible explanation).
The radio behaviour seems to correlate with position in the `Z'-shaped
track traced out on timescales of hours to days in the X-ray
colour-colour diagram (see Fig {\ref{ns4bw}}) in the sense that it is
strongest on the `horizontal branch' and weakest on the `flaring
branch', revealing an apparent anti-correlation with mass accretion
rate as in the black holes.

\begin{figure}
\centerline{\epsfig{file=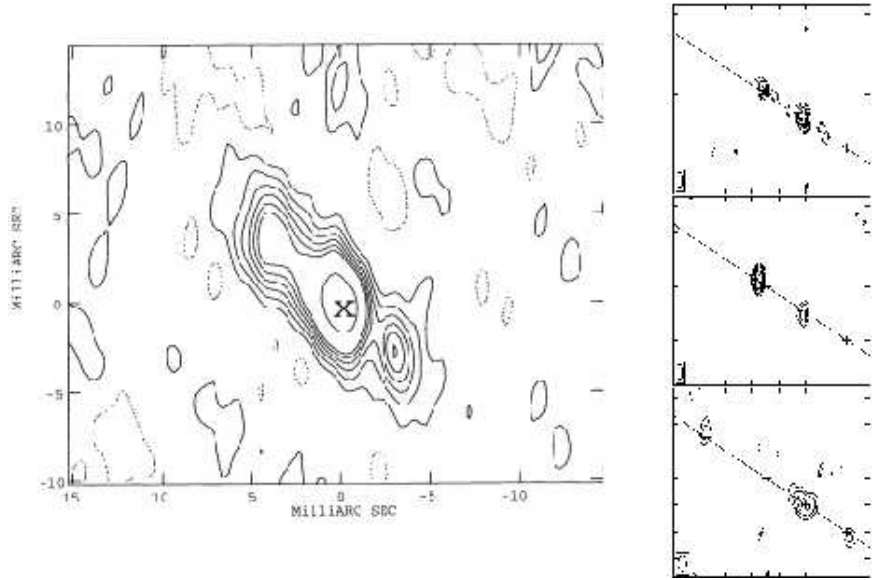,width=11.5cm,angle=0}}
\caption{(Left:) A VLBA image of milliarcsecond-scale radio jets from Sco X-1
with the core indicated by a `X'.
(from Bradshaw, Fomalont \& Geldzahler 1997).
(Right:) Multiple sequences of
such observations 
reveal
movement of the radio lobes at mildly relativistic velocities ($\sim
0.4c$) while being sporadically energised by a much more relativistic
($\Gamma \geq 2$) beam from the core (adapted from Fomalont et al. (2001a,b)).}
\label{fomalont}
\end{figure}

As noted above, intensive VLBI campaigns on Sco X-1 have revealed the
presence of a relativistic outflow (Bradshaw et al. 1999; Fomalont et
al. 2001a,b -- see Fig {\ref{fomalont}}). In particular, it seems that
following core radio flaring, relativistic ($\Gamma \geq 2$) beams are
acting on radio knots which themselves propagate away from the binary
core with mildly relativistic velocities. Given the similarity of the
radio properties between the six Z source, we can fairly confidently
conclude that they all have jets -- however, since the brightest
component is the core, it cannot yet be asserted that the jet-knot
interaction is occurring in all of them. Furthermore, caution should be
exercised in attempting to associate unresolved radio monitoring
(e.g. that performed in the 1990s with the Green Bank Interferometer)
with X-ray events (but see Hjellming et al. 1990b for a successful
experiment) since the delay between core events and subsequent
brightening in the knots is comparable to the timescale of motion in
the Z -- this may be an explanation for the `anomalous' observations
of GX 5-1 by Tan et al. (1991).

At present there is little study of, and consequently little evidence
for, possible extensions of the jet spectrum beyond the radio band in
the Z sources (although there are hints of some correlated optical
behaviour).  Estimates of the power in the jets are rather uncertain
(Fomalont et al. 2001b estimate super-Eddington power in the jets of
Sco X-1, but this is based upon the assumption that the major cooling
process is synchrotron losses, which is far from being clear), and are
at present based solely upon radio variability.

\subsection{`Atoll' sources}

It is worth re-stressing here that I am adopting a definition of
`atoll' source to mean all non-Z low magnetic field accreting neutron
stars -- this is considerably broader than the original definition of
Hasinger \& van der Klis (1989). Adopting this loose definition, atoll
sources are the single largest class of X-ray binary, contributing
around 45\% of the currently known population (in this classification,
`atoll' includes 'bursters', `dippers' etc). Investigation of their
disc-jet coupling, if any, is therefore of paramount interest -- not
least because they can exhibit `hard' X-ray states which are very
similar to the `low/hard' states of BHCs, while they of course remain
fundamentally different in possessing a solid surface.

Hjellming \& Han (1995) list the small number of reported radio
detections of atoll sources known at that time -- beyond some weak
detections of globular cluster sources, only GX 13+1 was repeatedly
detected at a relatively strong level (Garcia et al. 1988) -- in fact
at about the same radio luminosity as the Z sources (Fender \& Hendry
2000). However, GX 13+1 seems to be far from a `normal' atoll source
(Homan et al. 1998; Schnerr et al. 2003). It is interesting to note
that the three other brightest atoll sources, GX 3+1, GX 9+1 and GX
9+9 have never been detected in the radio band, and spend most of
their time in the soft `upper banana' (UB in Fig {\ref{ns4bw}}) state.

Most other atoll sources show somewhat harder spectra associated with
the `lower banana' (LB) or `island' (IS) X-ray states (which are
similar to the black hole `low hard' state).  Amongst these, Mart\'\i {}  et
al. (1998) reported repeated detections of the atoll source 4U 1728-34
(GX 354-0) at a level of up to $\sim 0.6$ mJy. In recent simultaneous
radio and X-ray observations of the same source, Migliari et
al. (2003) have revealed clear correlations between X-ray luminosity
and power spectral properties with the radio flux, establishing for
the first time a disc--jet coupling in such systems. Despite their
relative faintness in the radio band, it seems that there is a rich
phenomenology to be explored in these `hard' atoll sources.

\subsubsection{Neutron star transients}

There are a few detections of radio emission associated with neutron
star X-ray transients (see Fender \& Kuulkers 2001 for a list). These
include an unusual assortment of objects: the recurrent transient Aql
X-1 (Hjellming, Han \& Roussel-Dupre 1990), the first
accretion-powered msec pulsar SAX J1808.4-3658 (Gaensler, Stappers \&
Getts 1999) and 4U 1730-335 (`The Rapid Burster', Moore et al. 2000).
To hammer home a point made earlier, this author considers all these
sources to be quite similar in that they are low magnetic field
neutron stars accreting, on average, at a considerably sub-Eddington
rate, and I call them all `atoll' sources (it interesting to note that
there has not yet been a NS transient which displayed Z-type
properties even at the peak of outburst). The sample for NS transients
is considerably poorer than that for BH transients, something which
can be at least partially attributed to the fact that they are in
general fainter in the radio band (Fender \& Kuulkers 2001; see
section {\ref{bhvsns}}).

Cir X-1 can be considered as a recurrent neutron star transient
(perhaps comparable to GRS 1915+105 and Cyg X-3 in this respect); it
undergoes radio and X-ray flares every 16.6-days, during which periods
its X-ray luminosity is super-Eddington. This periodicity is
interpreted as heightened accretion during periastron passage of the
neutron star in a highly elliptical orbit -- essentially this system
undergoes repeated, periodic, soft X-ray transient outbursts. The
system is associated with an arcsec-scale one-sided radio jet (Fender
et al. 1998) embedded within an arcminute-scale radio nebula (Stewart
et al. 1993).  The X-ray classification of Cir X-1 has alternated
between `Z' and `Atoll' types, and at present it is not clear to which
category it belongs. 

\subsection{Black holes versus neutron stars}
\label{bhvsns}

There are clearly some broad similarities between black holes and
neutron stars in their X-ray : radio coupling. These include:

\begin{itemize}
\item{An association between states with hard X-ray spectra and strong
  X-ray variability and the presence of radio emission}
\item{An association between bright X-ray outbursts and radio flare
  events}
\item{In the brightest cases, the formation of large-scale radio lobes
  in the ISM}
\end{itemize}

\begin{figure}
\centerline{\epsfig{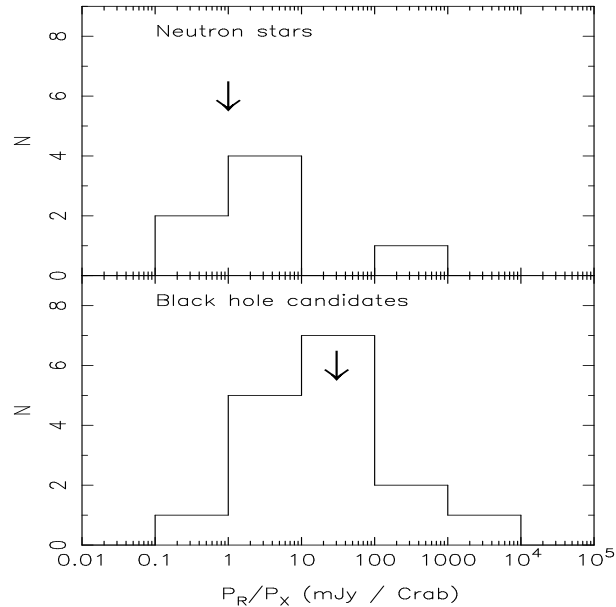}}
\caption{
Histograms of the `radio loudness' of neutron star (top) and black
hole (bottom) transients. The black holes are significantly more
`radio loud' than the neutron stars, by one to two orders of
magnitude. Also indicated by arrows are the mean `radio loudnesses' of
the neutron star Z sources and the brighter low/hard state black
holes, revealing the same trend. From Fender \& Kuulkers (2001).
}
\label{histo}
\end{figure}

Are there differences between jets from neutron stars and those from
black holes ?  There is at least one.  Fender \& Kuulkers (2001) have
found that defining a quantity `radio loudness' as the peak radio flux
of transients (in mJy), divided by their peak X-ray flux (in Crab),
{\em black hole transients are more radio loud than neutron stars}
(Fig {\ref{histo}}). Furthermore, by comparing the data for low/hard
state black holes and neutron star `Z' sources presented in Fender \&
Hendry (2000) they found a similar difference. In both classes of
object black holes seem to be about one or two orders of magnitude
more radio loud than neutron star systems. Migliari et al. (2003)
confirm a difference in radio luminosity by a factor $\sim 30$ between
the atoll source 4U 1728-34 in a hard state and low/hard state black
holes at a comparable Eddington ratio. This difference may be due to
greater photon (Compton) cooling of shocked electrons in the neutron
star systems, due to the presence of a radiating surface or low-level
magnetic field, or perhaps
due to some extra source of power in the black hole systems (Fender \&
Kuulkers 2001). A further possibility (Heinz \& Sunyaev 2003; Merloni,
Heinz \& di Matteo 2003; Falcke, K\"ording \& Markoff 2003) is that
the radio loudness scales with mass. However in this case, assuming the
`stellar mass' black holes are on average five times more massive than
the neutron stars, this suggests a rather steeper dependence on mass
than considered by these authors.

\begin{figure}
\centerline{\epsfig{file=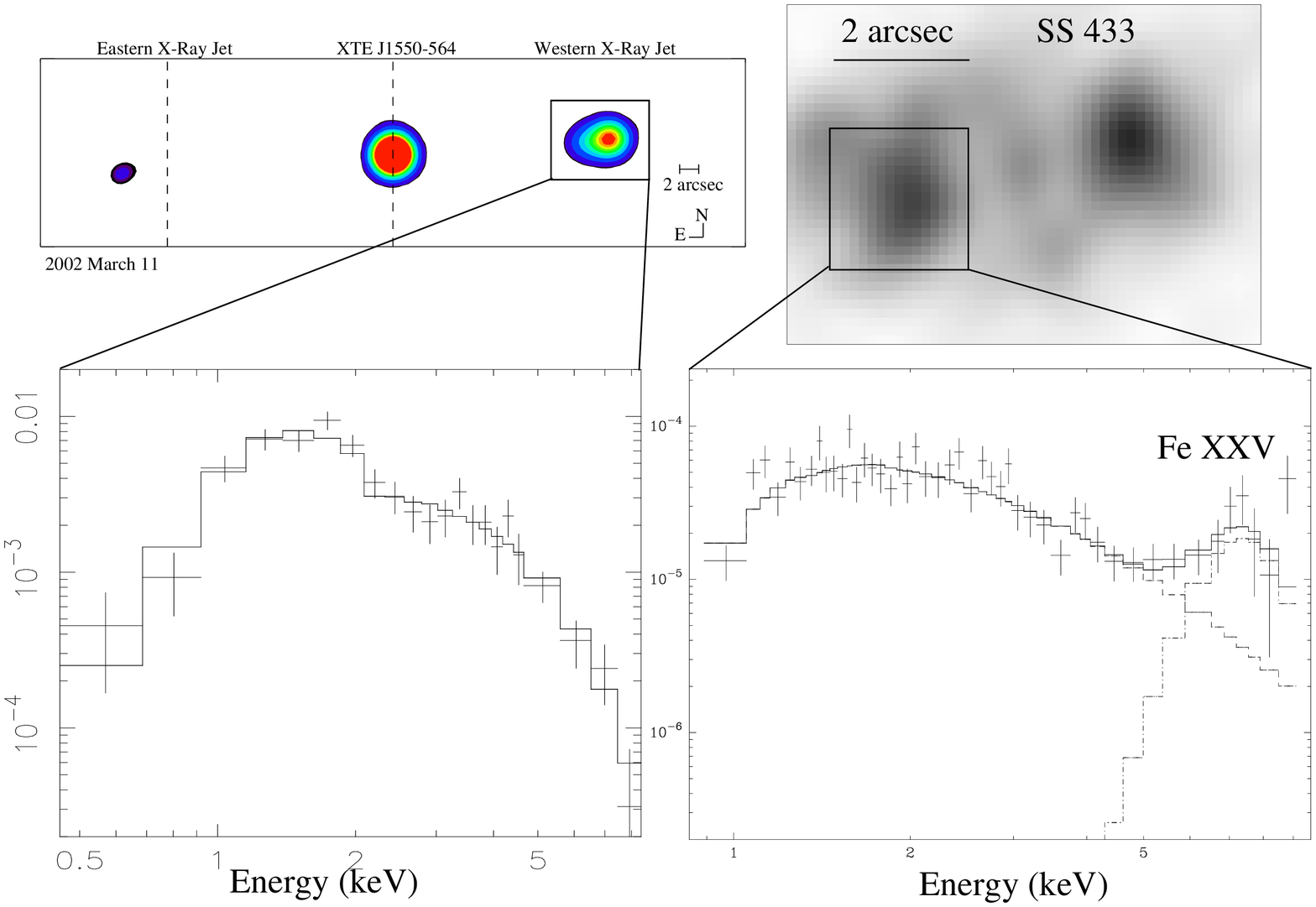,width=10cm,angle=270}}
\caption{Spatially resolved X-ray spectra of X-ray jets from the black
  hole transient XTE J1550-564 (left) and the persistent powerful jet
  source SS 433 (right). Note the strong emission line (probably Fe
  XXV) in the SS 433 spectrum, which is clearly not present in the
  jets of XTE J1550-564. Observations such as these demonstrate
  unequivocally that jets from X-ray binaries can be sources of both
  line-rich (`thermal') and featureless (`nonthermal') X-ray spectra
  which may be {\em beamed}. Adapted from Corbel et al. (2002), Kaaret et
  al. (2003), Migliari et al. (2002).}
\label{xspectra}
\end{figure}

\section{High energy / particle emission from jets}

Observations in the past two or three years have revealed
unambiguously that jets may be not only {\em associated} with phases
of high energy emission, but may actually be the {\em sites} of origin
of the observed emission.

\subsection{X-rays}

The possibility of some of the X-ray emission from X-ray binaries
arising in jets has already been alluded to in this text, and
explicitly suggested in the literature (e.g. Markoff, Falcke \& Fender
2001; Vadawale, Rao \& Chakrabarti 2001; Markoff et al. 2003; see also
Atoyan \& Aharonian 1999; Miller et al. 2002b). Before discussing this
further, it is worth restating the fact that in the past three years
{\em Chandra} imaging has unambiguously associated both thermal /
emission line (Migliari, Fender \& M\`endez 2002) and `hard' X-ray
spectra (Corbel et al. 2002; see also Angelini \& White 2003) with
jets from X-ray binaries (Fig {\ref{xspectra}}).

\begin{figure}
\centerline{\epsfig{file=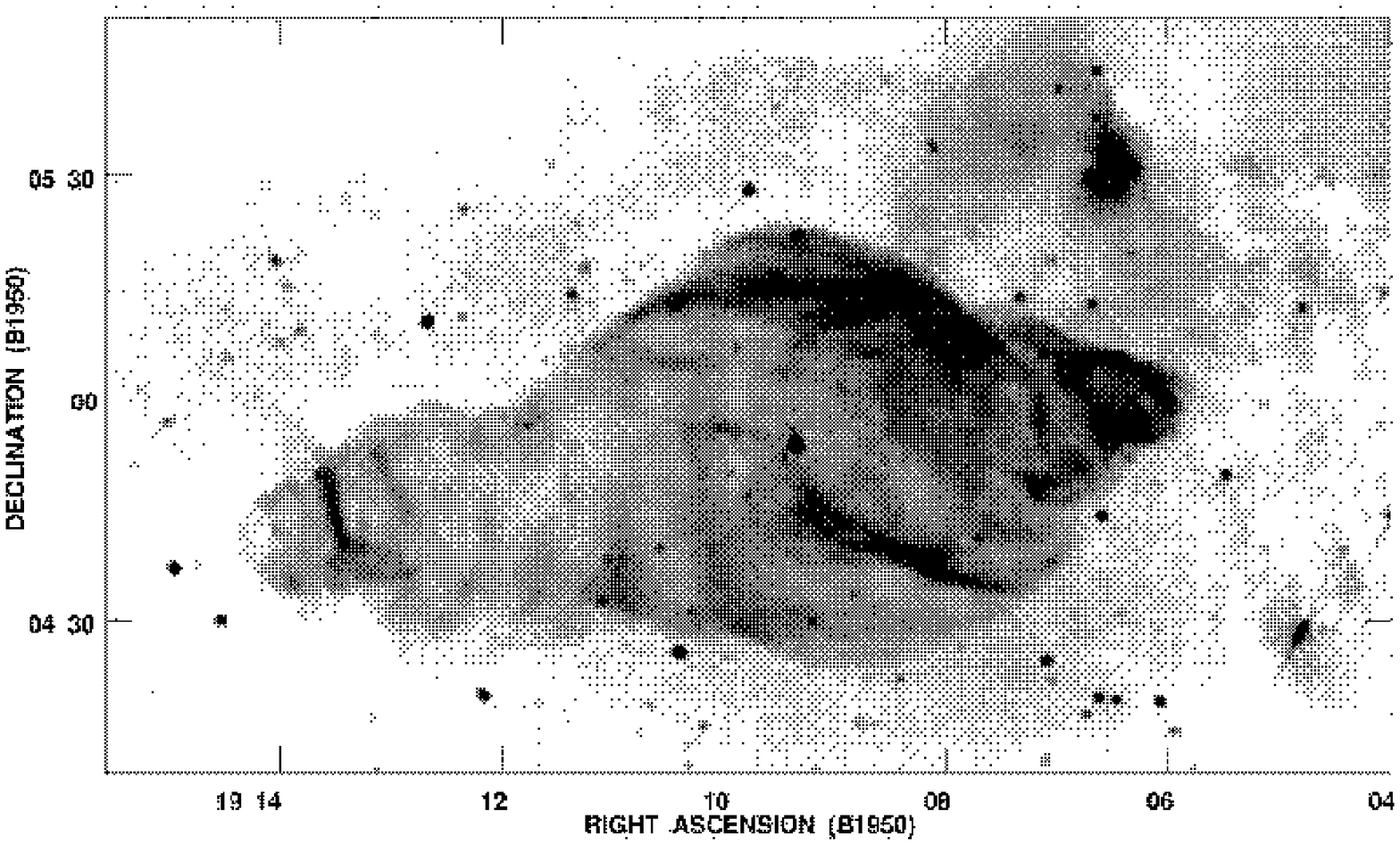,width=12cm,angle=0}}
\caption{The W50 nebula surrounding the powerful, quasi-continuous jet
source SS 433, which seems to have been distorted by the action of the
jets over thousands of years. From Dubner et al. (1998).}
\label{dubner}
\end{figure}

In a detailed model, Markoff et al. (2001, 2003; see Fig 9.8 and also
Falcke \& Biermann 1996, 1999) have suggested that the X-ray power law
observed in the low/hard X-ray state may in fact be the optically thin
synchrotron emission from the jet which is self-absorbed at lower
frequencies.  In fact as already noted a break from optically thick to
optically thin emission from the jet seems to have been found in the
right place for GX 339-4 (Corbel \& Fender 2002 -- in fact this may
have already been noted by Motch et al. 1985).  This is a radically
different interpretation for the origin of X-rays in this state, which
are generally ascribed to thermal Comptonisation (e.g. Poutanen 1998
and references therein; for more detailed objections to the model of
Markoff et al. see Zdziarski et al. 2003). The implication of the
model, if correct, would be that the majority of power output in the
low/hard state is in the form of a jet. Note that in this model the
X-ray emitting region would be spatially unresolvable and it is not
therefore an explanation for the extended X-ray jets observed from XTE
J1550-564 and SS 433 (Fig {\ref{xspectra}}). Vadawale, Rao \&
Chakrabarti (2001) have suggested that some component of the X-ray
spectrum of GRS 1915+105 may arise in synchrotron emission.
Georgonapoulos, Aharonian \& Kirk (2002) have suggested that X-ray
emission may originate due to Comptonisation by jet electrons of
photons from the companion star.

Returning to the large-scale X-ray jets, the fact that three have been
clearly imaged in the past $\sim 2.5$ years with Chandra (Marshall et
al. 2002; Migliari et al. 2002; Corbel et al. 2002; Angelini \& White
2003) indicates that they are likely to be rather ubiquitous. The fact
that X-ray emission, with a spectrum more or less indistinguishable
from the `off' state (for XTE J1550-564 and 4U 1755-33) may be
associated with beamed, long-lasting, jets is of considerable
interest. It certainly shows that jets, almost certainly via internal
or external shocks, may mimic faint hard states up to several years
after the binary source may have completely turned off. These are all
extra concerns for interpretations of the `quiescent' luminosities
of transient X-ray binaries: the X-ray jets of XTE J1550-564 (Corbel et
al. 2002; Kaaret et al. 2003; Tomsick et al. 2003) are more luminous
than most of the `quiescent' X-ray luminosities for black holes
reported in Garcia et al. (2001).

\subsection{High-energy / particle emission}

In an important recent work, Paredes et al. (2000; see also Rib\'o et
al. 2002 and Paredes et al. 2002) have reported a convincing
association between a massive X-ray binary with persistent radio jets
and an unidentified EGRET gamma-ray source. Their favoured scenario is
that relativistic electrons in the jet Comptonise photons from the
binary companion (similar to the model of Georgonapoulous et
al. 2002). The massive binary and probable jet source LS I +61 303
(Strickman et al. 1998; Gregory \& Neish 2002 and references therein)
may also be associated with a gamma-ray source, with a similar
physical origin a possibility.

Heinz \& Sunyaev (2002) have discussed the possible contribution of
X-ray binary jets to the production of galactic cosmic rays. They
conclude that, while in terms of overall energetics such jets are
still likely to inject less power into the ISM than supernovae, they
may contribute a specific and detectable component to the cosmic ray
spectrum. In particular, the shocks in the ISM associated with jets
from X-ray binaries will be considerably more relativistic than those
associated with the supernovae, and thus may be considerably more
efficient at particle acceleration.

Di Stefano et al. (2002) have suggested that jets from X-ray binaries
could be detectable sources of high-energy neutrinos. Kaiser \&
Hannikainen (2002) have further suggested that X-ray binary jets may
be the origin of a putative redshifted 511 keV annihilation line
observed from the direction of the X-ray transient GRS 1124-684
(however, an alternative explanation which is perhaps more widely
accepted is that the $\gamma$-ray emission was associated with a
transition of $^7$Li -- Martin et al. 1994). A possible explanation
for the $^7$Li production is spallation in the companion star
atmosphere due to the collision of a misaligned jet (Butt, Maccarone
\& Prantzos 2003).

In the light of the possibility of X-rays directly from jets, several
authors have considered the possibility of `micro-blazars' in which
jets aligned close to the line of sight could be observed as
$\gamma$-ray sources (e.g. Mirabel \& Rodriguez 1999; Kaufman Bernado
et al. 2002; Romero et al. 2002).

\section{Interactions}
\label{interaction}

As has already been alluded to in previous sections, it is becoming
clear that interactions between the jet, as launched by the
combination of accretion flow plus compact object, and the ambient
medium need to be taken into account for a full understanding both of
the radiation we observe and the internal physics of the outflows. Of
the classes of radio-emitting X-ray binaries only the weakest, the
atoll sources, have yet to provide us with a direct example of jet-ISM
interactions. These interactions have the potential to act as
independent measures of the power associated with jets from X-ray
binaries (`calorimeters'), although it has been argued that they may
be harder to detect than the corresponding lobes associated with AGN
(Heinz 2002; see also Levinson \& Blandford 1996). Furthermore, as
with AGN, it is possible that some of the presumed shock acceleration
may result not from jet-ISM interactions but from internal shocks
(Kaiser, Sunyaev \& Spruit 2000), perhaps resulting from varying flow
speeds (see also discussion in Migliari et al. 2002 for SS 433).

Considering first the black hole low/hard state sources, as well as
the milliarcsec-scale jet from Cyg X-1 (Stirling et al. 2001; Fig
{\ref{lowhard}}), arcmin-scale ($\equiv$ parsec-scale) jets have been
observed from 1E 1740-2942 and GRS 1758-258 (Mirabel et al. 1992;
Mart\'\i {}  et al. 2002; see Fig {\ref{17401758}}). It seems clear that
these larger lobes are the result of in-situ particle acceleration at
the interface between the steady jets and the ISM. 

Observations of large-scale radio and X-ray jets from the black hole
transient XTE J1550-564 (Corbel et al. 2002; Kaaret et al. 2003;
Tomsick et al. 2003) have provided us with unambiguous evidence of
broadband particle acceleration at the same time as the jet is
decelerating (Figs {\ref{corbel,decel}}).  Similarly, a one-sided
highly relativistic jet from Cyg X-3 on mas-scales (Mioduszewski et
al. 2001) seems to become a slower-moving, two-sided jet on
arcsec-scales (Mart\'\i {} , Paredes \& Peracaula 2001), indicating a
deceleration and in-situ particle acceleration. 

In the prototype of the Z sources, the brightest `persistent' neutron
stars, Sco X-1, Fomalont et al. (2001a,b) have found evidence for the
action of an unseen, highly relativistic flow, on radio-emitting
clouds which are themselves moving away from the binary core at mildly
relativistic velocities. The recurrent neutron-star transient Cir X-1
is associated not only with an asymmetric arcsec-scale radio jet but
with an arcmin-scale radio nebula (Stewart et al. 1993; Fender et
al. 1998) which, given the observed rapid timescale of radio
variability from this source, can be unambiguously associated with
in-situ particle acceleration, almost certainly powered by the jet.
The nebula around Cir X-1 provides a good example of the use of such
interaction zones as calorimeters -- a simple `minimum energy'
estimate indicates a total energy in the synchrotron emitting plasma
of $> 10^{48}$ erg, corresponding to, for example, three
thousands years' action at 1\% of the Eddington luminosity (see Heinz
2002 for further discussion). 

Most spectacularly, the persistent powerful binary jet source SS 433
powers the degree-scale W50 radio nebula (Fig {\ref{dubner}}), within
which are also located similar-scale X-ray jets (Brinkmann, Aschenbach
\& Kawai 1996). Note that on smaller arcsec scales there is already
evidence for reheating (Migliari et al. 2002), revealing that particle
re-acceleration is not only present, but occurs repeatedly at
different points in the flow.

In many, perhaps all, of these sources it now seems clear that a
picture of a single finite phase of particle acceleration following by
monotonic fading as the source expands and propagates away from its
launch site is far too simplistic. Multiple phases of particle
accelerations due to shocks -- whether internal or external -- are
perhaps instead the norm. While this necessarily complicates our
understanding of the disc--jet coupling (particularly when the various
physically distinct sites cannot be spatially resolved), it does, on
the other hand, allow us to constrain the power of jets in radio-quiet
phases such as the `high/soft' state. This follows because if these
states were producing powerful jets which for some reason
(e.g. extreme Compton cooling) were not radio-loud initially, we would
still expect the signatures of subsequent shock accelerations to be
found.

\section{Relation to other jet sources}

It is a common and useful exercise to compare accretion in X-ray
binaries with the analogous processes in related systems, most commonly
Cataclysmic Variables (CVs; see Kuulkers et al. this volume). In the
following I shall briefly compare X-ray binaries to other
jet-producing systems. Fig {\ref{meier}} indicates schematically possible
similarities and differences between jet formation in some of these
different classes of object.

\subsection{Active Galactic Nuclei}

The name `microquasar' (Mirabel et al. 1992) clearly reflects the
phenomenological similarities between jet-producing X-ray binaries and
Active Galactic Nuclei (AGN). Detailed quantitative comparisons are
only just beginning to be made; and will no doubt be the subject of
many future research papers. At the very roughest level, it is
tempting to associate the (disputed) `radio loud' and `radio quiet'
dichotomy observed in AGN with jet-producing (hard and transient) and
non-jet-producing (soft) states in X-ray binaries (see e.g. Maccarone,
Gallo \& Fender 2003).  Furthermore perhaps FRI jet sources can be
associated with the low/hard state and FRIIs with transients. Meier
(1999; 2001) has considered jet production mechanisms in both classes
of object, and drawn interesting parallels. Gallo, Fender \& Pooley
(2003; amongst others!) have made a qualitative comparison between
FRIs and low/hard state black hole X-ray binaries and FRIIs and
transients.

It is interesting to note that the short timescale disc-jet coupling
observed in GRS 1915+105 (Pooley \& Fender 1997; Eikenberry et
al. 1998; Mirabel et al. 1998; Klein-Wolt et al. 2001), in its most
basic sense -- that radio events are preceded by a `dip' and
associated spectral hardening in the X-ray light curve -- may also
have an analog in AGN: Marscher et al. (2002) have reported
qualitatively similar behaviour in 3C 120. 

Perhaps most exciting is the recent discovery that the power-law
relation between radio and X-ray luminosities found for low/hard state
BHCs (Corbel et al. 2001, 2003; Gallo, Fender \& Pooley 2003; see Fig
{\ref{gallo}}) may be directly relevant for the disc-jet coupling in
AGN. Merloni, Heinz \& di Matteo (2003) and Falcke, K\"ording \&
Markoff (2003) have both reported a `fundamental plane' of black hole
activity resulting from the coupling between mass, jet power and
accretion power. This plane matches almost perfectly with the Gallo,
Fender \& Pooley (2003) relation once the mass term is taken into
account, indicating truly similar physics across six to seven orders
of magnitude in mass. 

\subsection{Gamma ray bursts}

While currently observations allow that X-ray binary jets may achieve
on occasion bulk Lorentz factors as large as those of the fastest AGN
jets (Fender 2003), Gamma ray bursts (GRBs) appear to belong to
another regime, with $\Gamma > 100$ (e.g. Baring \& Harding 1997;
Lithwick \& Sari, 2001). While the physics of jet interaction and
emission may be similar, being based upon shock acceleration and the
synchrotron process, the workings of the jet-producing engine in GRBs
are so buried that it is hard to know how to make quantitative
comparisons. Nevertheless, such comparisons should be attempted, and
the differences between XRB transients, some of which reach
super-Eddington rates, and GRBs, may not be as great as currently
thought. Since in X-ray binaries we are fairly confident that to some
degree the jet activity reflects that in the accretion flow, it may be
conceivable that the (highly compressed) patterns of behaviour in GRBs
(originating in the jet) may reveal similarities with the slower black
hole accretion processes observed in XRBs.

\subsection{Other galactic jet sources}

X-ray binaries aside, there are multiple other sources of jets
associated with `stellar'-scale objects within our galaxy (and
presumably others). However, in no other class of sources are there
truly relativistic jets associated with accretion.

There are however nonrelativistic jets associated with accretion in
(at least) Young Stellar Objects (YSOs; e.g. Lada 1985; Reipurth \&
Bally 2001) and Super Soft Sources (SSS; see Kahabka in this volume
for a full description). The SSS can perhaps be most clearly compared
to the X-ray binaries since they seem to be producing highly
collimated jets as a result of high accretion rates in a binary
(Cowley et al. 1998 and references therein), albeit at much lower
velocities ($0.01c$ or less). These jets are revealed not by their
radio emission but by optical twin optical/infrared lines originating
from the jets (reminiscent of SS 433).  The `symbiotic' binary CH Cyg
is another interesting source of sub-relativistic jets associated with
accretion. These jets {\em do} emit in the radio band, and may be
precessing (Crocker et al. 2002); furthermore Sokoloski \& Kenyon
(2003) have reported a possible disc-jet coupling similar to that
found in GRS 1915+105.  Finally it is often noted that radio pulsars
such as the Crab and Vela seem to be associated with (relativistic)
jets and yet are not accreting (e.g. Blandford 2002).

One conclusion has been drawn from the comparison of X-ray binaries
with such diverse galactic objects: that the jet velocity is always
comparable to the escape velocity of the accreting object (e.g. Livio
1999; Mirabel \& Rodr\'\i guez 1999). While this seems to hold over the
sub-relativistic and mildly-relativistic regime, evidence for varying
jet speeds from the same black hole, and for $\Gamma > 2$ flows from
neutron stars seem to indicate that it is not a hypothesis which can
be extrapolated beyond neutron stars.

\subsection{Ultraluminous X-ray sources}

Ultraluminous X-ray sources are X-ray sources in external galaxies
with apparent isotropic luminosities requiring black hole masses of
$\sim 100 M_{\odot}$ or more in order to remain sub-Eddington (i.e. at
least a factor of a few more luminous that GRS 1915+105). There are at
present three competing explanations for these sources, all involving
accretion onto a black hole. If the radiation really is isotropic then
`intermediate mass black holes' are invoked (e.g. Colbert \& Mushotzky
1999); alternatively the radiation may be anisotropically emitted from
the accreting region (King et al. 2001) or relativistically aberrated
due to e.g. an origin in a jet (K\"ording, Falcke \& Markoff 2002; see
also Georganopoulos et al. 2002). At the moment the nature of ULXs
remains unclear (see more detailed discussion by King, this volume).

An obvious prediction of the jet model would be radio counterparts to
such sources, and there is tantalising evidence that this may have
recently been achieved.  Dubus \& Rutledge (2002) have suggested that
the X-ray source M33 X-8 may be associated with a weak radio source;
Kaaret et al. (2003) claim to have identified the radio counterpart to
an ULX in NGC 5408. While these claims need confirmation, observations
of the radio counterparts of such sources will surely provide strong
clues to their intrinsic nature.

\section{On the origin of jets}

\begin{figure}
\centerline{\epsfig{file=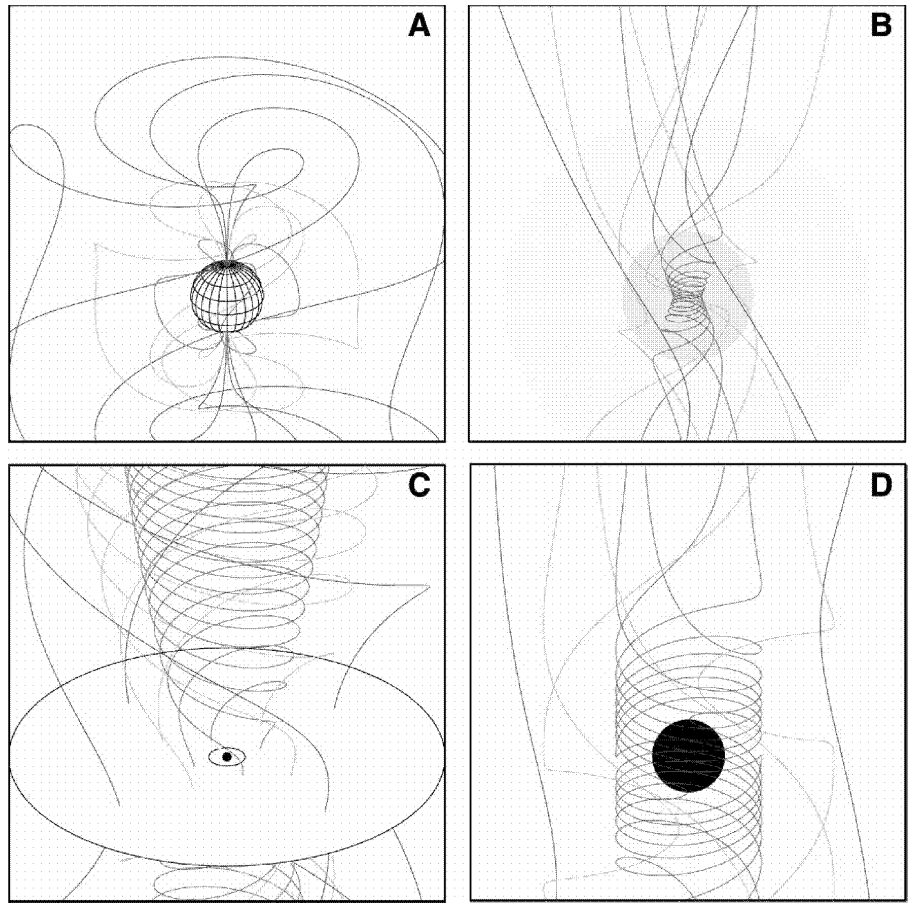,width=12cm,angle=0}}
\caption{Four ways to make jets with magnetic fields. {\bf A:} dipole
  field of a rotating neutron star. {\bf B:} A collapsing object
  drawing and winding up an initially uniform field. {\bf C:} Poloidal
  magnetic field from a magnetised accretion disc. {\bf D:}
  Frame-dragging near a rotating black hole resulting in strong
  coiling of the magnetic field lines. Types {\bf C} and {\bf D},
  although possibly also {\bf A} may be relevant for X-ray binaries;
  type {\bf A} for isolated pulsars; types {\bf C} and {\bf D} for
  AGN, and types {\bf B},{\bf C} or {\bf D} may be relevant for gamma
  ray bursts. From Meier, Koide \& Uchida (2001).}
\label{meier}
\end{figure}

In this review the observational properties of jets from X-ray
binaries have been considered and some broad-ranging empirical
relations have been established (most notably the association of jets
with `hard' X-ray states). Such empirical connections require
theoretical interpretation and the theory community has in recent
years begun to rise to the task (motivated at least in part by a
desire to use X-ray binary jets to explain those of AGN). There is
certainly no room here to discuss these theoretical developments in
detail, but it is worth pointing out some key relevant works.

Blandford \& Payne (1982; see also e.g. Ogilvie \& Livio 2001)
provided the groundwork for models in which magnetic fields rooted in
an accretion flow may produce `radio' jets.  The association of the
`low/hard' state with $\sim$steady jet formation has been interpreted
by Meier (2001) and Meier, Koide \& Uchida (2001) as strong evidence
for MHD jet formation. In this scenario the strongest jets result from
accretion flows with a large scale height, and so the jets are
naturally suppressed in `high/soft' accretion states which are
dominated by a geometrically thin accretion disc.  Merloni \& Fabian
(2002) discuss `coronal outflow dominated accretion discs' in which
they balance both accretion and outflow powers.  Livio, Pringle \&
King (2003) have put forward a model in which the hard X-ray states of
BHCs represent modes in which the bulk of the accretion energy is
deposited into the bulk flow of a relativistic jet. Such `jet
dominated' states may be empirically borne out by observations
(Fender, Gallo \& Jonker 2003). Lynden-Bell (2003) discusses the
formation by magnetised accretion discs of `towers' which can
collimate jets. In all these theoretical models a magnetised accretion
flow is the basis of a MHD outflow; given the widespread acceptance of
the magneto-rotational instability (MRI) as the origin of accretion
disc viscosity (e.g. Balbus \& Hawley 1991; Turner, Stone \& Sano
2002) this highlights the probably key role of magnetic fields in the
coupled accretion -- outflow system (see e.g. Kudoh, Matsumoto \&
Shibata 2002 for a discussion of a possible relation between MRI and
jet formation). Fig {\ref{meier}} presents four different
configurations of accretion with magnetic fields which may result in
jet formation.
In a rather different but still
magnetically-orientated approach Tagger \& Pellat (1999; see also
e.g. Varniere \& Tagger 2002), in the `accretion-ejection instability'
model have suggested that an instability related to the vertical
component of magnetic field in the inner regions of accretion discs
may result in the transport of energy and angular momentum away from
the accretion flow, possibly powering a jet or wind. In this, and the
related works of Das, Rao \& Vadawale (2003) and Nobili (2003), the
jet should be intimately coupling to the timing properties of the
accretor (of course this {\em is} already empirically observed to a
certain extent since the different states of both BHCs and NS X-ray
binaries have different timing properties).

As an alternative to magnetic acceleration, radiative acceleration
(e.g. the `Compton Rocket' of O`Dell 1981) is unlikely to be able to
push jets to the highest observed bulk velocities (Phinney 1982) but
may still be operating, via line-locking, in the case of SS 433
(Shapiro, Milgrom \& Rees 1986).

Many variants on radiatively inefficient accretion flows are now
beginning to consider outflows as part of their solutions
(e.g. Narayan \& Yi 1995; Blandford \& Begelman 1999; Das 1999;
Beckert 2000; Markoff et al. 2001; Becker, Subramanian \& Kazanas
2001). It remains to be seen which, if any, of these models comes
closest to reproducing the observational characteristics of accretion
onto black holes at a variety of rates, but note that numerical
simulations of radiatively inefficient accretion flows also seem to
form jets and outflows (Hawley \& Balbus 2002).  In fact already more
than two decades ago Rees et al. (1982) discussed a likely connection
between `ion-supported tori' (essentially advective flows) and the
formation of radio jets. While Rees et al. (1982) were motivated by
the study of AGN, they noted that `..relativistic jets collimated by
tori around stellar-mass black holes may exist within our Galaxy.'.

Are we ever going to be able to directly image the jet formation
region in X-ray binaries ? It seems unlikely -- Junor, Biretta \&
Livio (1999) report the direct imaging of jet formation around the
(low-luminosity) AGN M87 at a distance of $\sim 100$ Schwarzschild
radii from the black hole. Comparing M87 to X-ray binaries in our own
galaxy, the ratio of distances is so much smaller than the ratio of
black hole masses (and therefore Schwarzshild radii), that such
imaging will not be possible. For example, a structure of size 100
Schwarschild radii around a $10 M_{\odot}$ black hole at a distance of
5 kpc would have an angular size of $\sim 10^{-11}$ arcsec ! Therefore
the key to studying jet {\em formation} in X-ray binaries will remain
in careful multiwavelength studies at the highest time resolution,
such as those performed with such success on GRS 1915+105
(e.g. Mirabel et al. 1998; Klein-Wolt et al. 2002).

\subsection{On black hole spin}

It has been suggested both for AGN and for XRBs that the jets may in
whole or in part be powered by the spin of the black hole
(e.g. Blandford \& Znajek 1977; Koide et al. 2002), although Livio,
Ogilvie \& Pringle (1999) argue that the energy extracted from the
black hole in this way will never exceed that from the inner regions
of the accretion disc.  For the black holes in the low/hard state the
apparently tight and universal correlation between X-ray and radio
fluxes seems to indicate that either:

\begin{itemize}
\item{Black hole spin {\em is not} important for the formation of jets in
  the low/hard state of black holes. This may be natural if the jets are
  formed at large distances from the black hole.}
\item{Black hole spin {\em is} important, and all the binary black
  holes have about the same (dimensionless) spin parameter. This may
  be natural since they all originate in rotating massive stars
  (c.f. radio pulsars).}
\end{itemize}

In this context it may well be the case that even if black hole spin
is not important for the low/hard state, it may well still be for the
(transient) relativistic ejections which show a much greater scatter
(although this may also be attributed to stronger beaming and less
comprehensive coverage of light curves -- Fender \& Kuulkers 2001;
Gallo, Fender \& Pooley 2003).  Furthermore, it should be noted that
these conclusions are rather contrary to those of Cui, Zhang \& Chen
(1998) who conclude that (a) most binary black holes are only slowly
spinning, (b) only rapidly spinning black holes produce radio jets.

\section{Conclusions}

In this review I have attempted to summarise our observational
understanding of the phenomena of jets from X-ray binaries. In the
process I have lightly, but no more, touched on various
interpretations currently at large in the literature. 

It is interesting to note that, whereas they were poorly investigated
or understood one decade ago, these jets are now being considered as
possible explanations for many exotic or high energy phenomena. Not
only do they clearly emit from 100s of MHz to at least several keV, a
range of $10^{10}$ in photon energy, but they may be important sites
of particle acceleration in the ISM and even sources of neutrinos. One
thing seems clear - they are {\em powerful} and need to be carefully
considered when attempting to describe the physics of accretion onto
compact objects. This author has no doubt that the next decade will
provide yet more excitement and surprises in this field.

\section*{Acknowledgements}

The author would like to thank Catherine Brocksopp, Stephane Corbel,
Elena Gallo, Sebastian Heinz, Thomas Maccarone, Sera Markoff and
Simone Migliari for a careful reading of the manuscript and numerous
useful suggestions.

\end{document}